\documentclass[sigconf]{acmart}%authordraft
\AtBeginDocument{%
  \providecommand\BibTeX{{%
    \normalfont B\kern-0.5em{\scshape i\kern-0.25em b}\kern-0.8em\TeX}}}

\copyrightyear{2021}
\acmYear{2021}
\setcopyright{acmlicensed}\acmConference[SIGIR '21]{Proceedings of the 44th
International ACM SIGIR Conference on Research and Development in
Information Retrieval}{July 11--15, 2021}{Virtual Event, Canada}
\acmBooktitle{Proceedings of the 44th International ACM SIGIR Conference on
Research and Development in Information Retrieval (SIGIR '21), July 11--15,
2021, Virtual Event, Canada}
\acmPrice{15.00}
\acmDOI{10.1145/3404835.3462875}
\acmISBN{978-1-4503-8037-9/21/07}

\newcommand{\ie}{\emph{i.e., }}
\newcommand{\eg}{\emph{e.g., }}

\newcommand{\cf}{\emph{cf. }}

\usepackage{enumitem}
\usepackage{verbatim}
\usepackage[ruled]{algorithm2e}
\usepackage{multirow}
\usepackage{subfigure} 
\usepackage{ragged2e}
\usepackage{caption}
\usepackage{graphicx}
\usepackage[normalem]{ulem}
\useunder{\uline}{\ul}{}
\usepackage{amsmath}
\usepackage{setspace}
\begin{document}

%%
%% The "title" command has an optional parameter,
%% allowing the author to define a "short title" to be used in page headers.
% \title{Popular is Unequal to Interest: Deconfounding Item Popularity in Recommender System}
\fancyhead{}
\title{Causal Intervention for Leveraging Popularity Bias in Recommendation}
% \renewcommand{\title}{Causal Intervention for Leveraging Popularity Bias in Recommendation}
%%
%% The "author" command and its associated commands are used to define
%% the authors and their affiliations.
%% Of note is the shared affiliation of the first two authors, and the
%% "authornote" and "authornotemark" commands
%% used to denote shared contribution to the research.

\author{Yang Zhang$^{1\dag}$, Fuli Feng$^{2*}$, Xiangnan He$^{1*}$, Tianxin Wei$^1$, Chonggang Song$^3$, Guohui Ling$^3$ and Yongdong Zhang$^1$}

\def\authors{Yang Zhang, Fuli Feng, Xiangnan He, Tianxin Wei, Chonggang Song, Guohui Ling, Yongdong Zhang}

\affiliation{\institution{$^1$University of Science and Technology of China,$^2$National University of Singapore,$^3$Tencent Inc.\country{}}}

% \affiliation{
% \institution{$^3$WeChat, Tencent\city{Shenzhen}\country{China}}}

\email{zy2015@mail.ustc.edu.cn,{fulifeng93,xiangnanhe}@gmail.com,rouseau@mail.ustc.edu.cn}
\email{{jerrycgsong,randyling}@tencent.com,zhyd73@ustc.edu.cn}

% \thanks{$\dag$This work is supported by the National Natural Science Foundation of China (U19A2079, 61972372) and National Key Research and Development Program of China (2020AAA0106000).$*$ Corresponding authors. $\ddag$ Yang Zhang did this work at Tencent.  }
\thanks{* Corresponding author. }
\thanks{$\dag$ Work done at Tencent.}
%$*$ Fuli Feng and Xiangnan He are the co-corresponding authors. $\ddag$Yang Zhang did this work when he was an intern at WeChat, Tencent
% \thanks{$*$Fuli Feng and Xiangnan He are the co-corresponding authors. $\ddag$Yang Zhang did this work when he was an intern at WeChat, Tencent}
% \thanks{}
% \footnotetext[1]{You can add acknowledgements here.}

% \author{paper id:361}
%\author{xx}
%\affiliation{%
%  \institution{xxxxxx}
%   \streetaddress{1 Th{\o}rv{\"a}ld Circle}
%   \city{Hekla}
%  \country{xxx}}
%\email{xxxx}

%%
%% By default, the full list of authors will be used in the page
%% headers. Often, this list is too long, and will overlap
%% other information printed in the page headers. This command allows
%% the author to define a more concise list
%% of authors' names for this purpose.
% \renewcommand{\shortauthors}{anonymous}

%%
%% The abstract is a short summary of the work to be presented in the
%% article.
\begin{abstract}

Recommender system usually faces popularity bias issues: from the data perspective, items exhibit uneven (usually long-tail) distribution on the interaction frequency; from the method perspective, collaborative filtering methods are prone to amplify the bias by over-recommending popular items. It is undoubtedly critical to consider popularity bias in recommender systems, and existing work mainly eliminates the bias effect with propensity-based unbiased learning or causal embeddings. However, we argue that not all biases in the data are bad, \ie some items demonstrate higher popularity because of their better intrinsic quality. Blindly pursuing unbiased learning may remove the beneficial patterns in the data, degrading the recommendation accuracy and user satisfaction. 

This work studies an unexplored problem in recommendation --- how to \textit{leverage} popularity bias to improve the recommendation accuracy. The key lies in two aspects: how to remove the bad impact of popularity bias during training, and how to inject the desired popularity bias in the inference stage that generates top-$K$ recommendations. 
%intervene popularity distribution during inference since popularity is always dynamically changing. 
This questions the causal mechanism of the recommendation generation process. Along this line, we find that item popularity plays the role of \textit{confounder} between the exposed items and the observed interactions, causing the bad effect of bias amplification. To achieve our goal, we propose a new training and inference paradigm for recommendation named
\textit{Popularity-bias Deconfounding and Adjusting} (PDA). It removes the confounding popularity bias in model training and adjusts the recommendation score with desired popularity bias via causal intervention. We demonstrate the new paradigm on the latent factor model and perform extensive experiments on three real-world datasets from Kwai, Douban, and Tencent. Empirical studies validate that the deconfounded training is helpful to discover user real interests and the inference adjustment with popularity bias could further improve the recommendation accuracy. We release our code at https://github.com/zyang1580/PDA.%The code is available at https://github.com/zyang1580/PDA. 

\end{abstract}

%%
%% The code below is generated by the tool at http://dl.acm.org/ccs.cfm.
%% Please copy and paste the code instead of the example below.
%%


%%
%% Keywords. The author(s) should pick words that accurately describe
%% the work being presented. Separate the keywords with commas.

%% A "teaser" image appears between the author and affiliation
%% information and the body of the document, and typically spans the
%% page.

% \begin{teaserfigure}
%   \includegraphics[width=\textwidth]{sampleteaser}
%   \caption{Seattle Mariners at Spring Training, 2010.}
%   \Description{Enjoying the baseball game from the third-base
%   seats. Ichiro Suzuki preparing to bat.}
%   \label{fig:teaser}
% \end{teaserfigure}

%%
%% This command processes the author and affiliation and title
%% information and builds the first part of the formatted document.
\maketitle
\begin{spacing}{0.936}
\section{Introduction} \label{sec:intro}

%% 1. popularity bias and importance
Recommender system provides personalized service for users to seek information, playing an increasingly important role in a wide range of online applications, such as e-commerce, news portal, content-sharing platform, and social media. %plays an important role in tackling the information explosion of the online world, such as e-commerce \cite{EchoChambers}, advertising \cite{DIN} and video-sharing social networks \cite{youtubeNet}. 
However, the system faces popularity bias issues, which stand on the opposite of personalization. %usually suffers from the widely existed popularity bias. 
%Generally, popularity bias is described in two aspects. 
%On the one hand, from the data perspective, few popular items occupy the most of interactions but the majority of other items receive little attention, as it's well known as the frequency of items of interaction has a long-tail shape \cite{long_tail, Himan_reg}. 
On one hand, the user-item interaction data usually exhibits long-tail distribution on item popularity --- a few head items occupy most of the interactions whereas the majority of items receive relatively little attention~\cite{FeedbackLoop, Himan_reg}. 
%from the data perspective, the interaction frequency of item is a typically long-tail distribution, where few popular items occupy the most of interactions but the majority of other items receive little attention~\cite{FeedbackLoop, Himan_reg}. 
%On the other hand, from the model perspective, due to the uneven distribution of data and the drawbacks of learning method and model self, the model usually does not only inherit the bias in data but also amplify the bias, lead to the popular get more popular and the unpopular become more unpopular \cite{Himan}. 
On the other hand, the recommender model trained on such long-tail data not only inherits the bias, but also amplifies the bias, making the popular items dominate the top recommendations~\cite{UMAP_recbias,Himan-thesis}.
% What's worse, with the closed feedback loop of the systems \cite{FeedbackLoop}, the bias will cause many undesirable phenomenons, such as echo chamber \cite{EchoChambers}, filter bubble \cite{Calibrated}  and Matthew effect \cite{Mattheweffect}. 
%Worse still, with the closed feedback loop of the recommender system~\cite{FeedbackLoop}, the bias will cause many undesirable phenomenons, such as echo chamber \cite{EchoChambers}, filter bubble \cite{Calibrated} and Matthew effect \cite{Mattheweffect}, damaging the interests of stackholders in the system, including the user, content provider, and the platform~\cite{System-Level,Himan-thesis}.
Worse still, the feedback loop ecology of recommender system further intensifies such Matthew effect~\cite{Mattheweffect}, causing notorious issues like echo chamber~\cite{EchoChambers} and filter bubble~\cite{Calibrated}.
%And the impacts of the popularity bias will be paid to multi-stakeholders \cite{Himan_multi}, such as users, content providers, and the systems selves. % For items, some items don't get the enough recommended chances that they should have; For users,  some users are recommended more popular items that don't match their interest; For systems, the dissatisfactions of both providers of items and users will bring the various losses. 
Therefore, it is essential to consider the popularity bias issue in recommender systems.
%So it is necessary to consider popularity bias in the recommendation system. 

%% previous works
%Before our work, there are some works try to deal with the popularity bias. They can be split into three categories by their technical approaches:
%The existing work attempts to achieve the target by pursuing unbiased recommendation, including:
Existing work on popularity bias-aware recommendation mainly performs unbiased learning or ranking adjustment, which can be categorized into: 
\begin{itemize}[leftmargin=*]
    %\item Inverse Propensity Scoring (IPS, also named Inverse Propensity Weighting) \cite{IPS_rec}, which try to get unbiased effect estimation by adjusting the data distribution to be even with paying higher weights for clicks on unpopular items. Ideally, it can get an unbiased estimation if the gotten distribution doesn't have any biases. But usually, the propensity scores are hard to be accurately estimated and high variances of the scores will influence the effectiveness.  
    \item \textit{Inverse Propensity Scoring} (IPS), which adjusts the data distribution to be even by reweighting the interaction examples for model training~\cite{IPS_rec,rec_ips_cap_norm}. Although IPS methods have sound theoretical foundations, they hardly work well in practice due to the difficulties in estimating propensities and high model variance. 
    %increasing the weight of interactions on unpopular items during model training. 
    %The IPS method has well founded theoretical support, however, faces issue of inaccurate estimation and high variances of the propsensity in practice.
    %It's a widely used method to estimate causal effects by adjusting the data distribution to a target distribution in which correlation is equal to causality in a re-weighting way with the propensity scores. Ideally, it can get an unbiased estimation. But usually, the propensity score is hard to estimate, and the propensity scores are vulnerable to noise and have high variance. The drawbacks will significantly affect the performance. In recommendation, it imposes lower weights for clicks on popular items and boosts long-tail items\cite{IPS_rec1,IPS_rec2,IPS_rec3}, i.e. the target thought as an even distribution. According to previous work, \cite{DICE}, it has poor performance.      
    %\item \textit{Causal Embedding}, which try to use even data \cite{2018causalE, DICE} or extra \cite{DICE} hand-defined "causal-specific" data to guide models to obtain unbiased embedding in the framework of multi-tasks or curriculum learning, where the causal-specific data usually emphasize long-tail items. The performances rely on the referenced data.
    \item \textit{Causal Embedding}, which uses bias-free uniform data to guide the model to learn unbiased embedding~\cite{2018causalE,distillation}, forcing the model to discard item popularity. However, obtaining such uniform data needs to randomly expose items to users, which has the risk of hurting user experience. Thus the data is usually of a small scale, making the learning less stable. 

    \item \textit{Ranking Adjustment}, which performs post-hoc re-ranking on the recommendation list~\cite{xquad2019,BPR_PC} or model regularization on the training~\cite{Himan_reg,BPR_PC}. 
    %~\cite{xquad2019,BPR_PC,Himan_. reg,UMAP_recbias}
    Both types of methods are heuristically designed to intentionally increase the scores of less popular items, which however lack theoretical foundations for effectiveness. 
    %. Tless pophere are two types of methods. The first is post-hoc re-ranking, which only re-ranks recommendation lists without changing the learning process. The second is the regularization-based method which adds an extra term in the loss to increase the probabilities of unpopular items to get high scores. %which restricts the difference of predicted scores between popular and unpopular items in training.%  % which constrainedly  decreases the correlation between popularity and predicted scores.
    %Both of them are heuristic designed and have not theoretical foundations for their effectiveness.
    %\item Reranking \cite{xquad2019,BPR_PC} and regularization \cite{ControlPOPreg17,BPR_PC,UMAP_recbias}. Both of the two methods try to lift the rate of the unpopular item in the recommendation list. Re-ranking is a post-processing method, it doesn't affect the process of learning. So it can be applied to any recommender system to adjust the recommendation list. But the interests referenced when re-ranking are still biased. Regularization realizes the same goal when training by adding an extra term to loss to make the prediction fair for popular and unpopular items. \cite{BPR_PC} show that re-ranking methods have better performance. 
\end{itemize}

%All the strategies try to eliminate the bias effect, in especial, some methods take even status as target status. However, we argue that leveraging popularity bias can improve the recommendation accuracy. 
Instead of eliminating the effect of popularity bias, we argue that the recommender system should \textbf{leverage} the popularity bias. The consideration is that not all popularity biases in the data mean bad effect. For example, 
%At first, not all biases in the data are bad. The key reason is obvious --- 
some items demonstrate higher popularity because of better intrinsic quality or representing current trends, which deserve more recommendations. %Removing all the bias without distinguishing, especially when expecting a uniform state, will undoubtedly result in the loss of some very effective patterns in data, making intrinsically high-quality items can't acquire enough opportunity, that they should have, to be recommended. 
Blindly eliminating the effect of popular bias will lose some important signals implied in the data, improperly suppressing the high-quality or fashionable items. %undoubtedly result in the loss of some very effective preference signal, making the items of high-quality receive less opportunities than they deserve. 
%What's more, in some cases, introducing extra popularity bias is also desired, such as we can predict which items will be hotter than others with the same quality in one period, the more hot items are also worth to be recommended more than the others in the period. 
Moreover, some platforms have the need of introducing desired bias into the system, \eg promoting the items that have the potential to be popular in the future. 
% As such, leveraging popularity bias is worth to be expected. But how to remove the other bad effect of popularity bias and how to inject the desired popularity is hard to answer if we don't know where the bad effects come from and What role popularity plays.
This work aims to fill the research gap of effectively leveraging popularity bias to improve the recommendation accuracy. 
%In this work, we study the central theme of leveraging popularity bias in recommender system, where challenges are twofold: 1) removing the bad effect of popularity bias for model training; and 2) injecting the desired popularity bias into the recommendations.
%Fortunately, the causal story behind the recommendation generation process can answer the latter two questions.  

\begin{figure}
\centering
\subfigure[Causal graph of traditional methods.]{\label{fig:a}\includegraphics[width=0.15\textwidth]{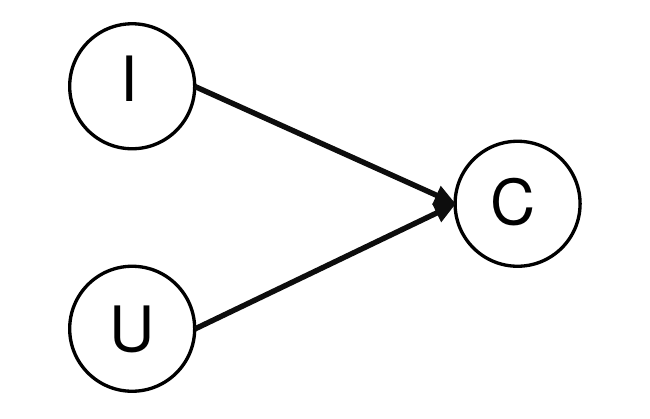}}
\subfigure[Causal graph that considers item popularity.]{\label{fig:b} \includegraphics[width=0.14\textwidth]{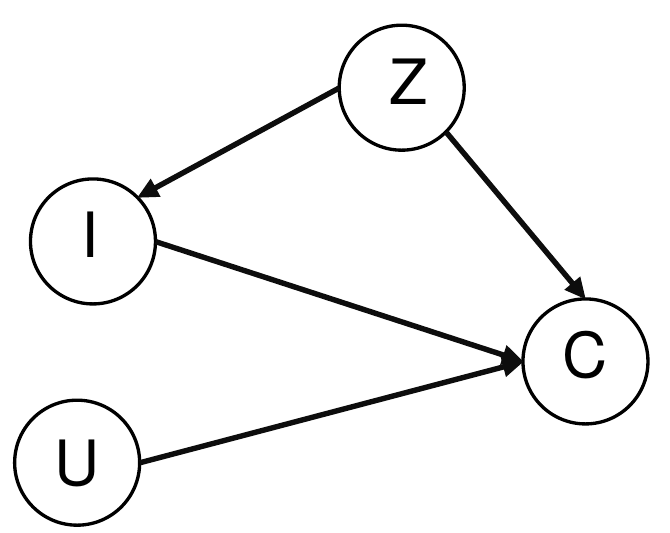}}
\subfigure[We cut off $Z\rightarrow I$ for model training.]{\label{fig:c}\includegraphics[width=0.13\textwidth]{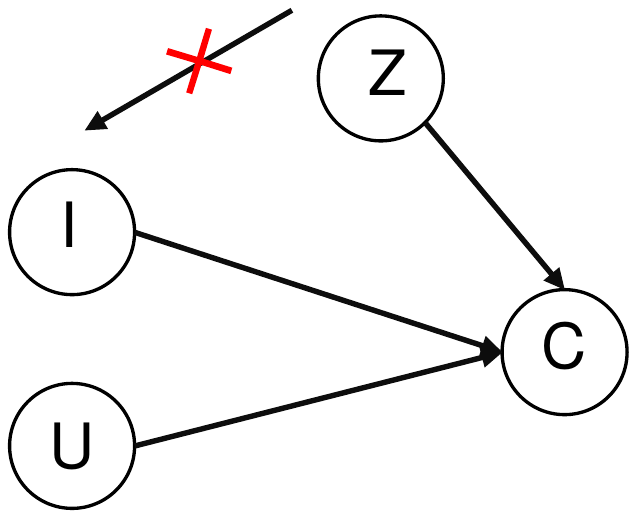}}
\vspace{-10pt}
\caption{Causal graphs to describe the recommendation process. U: user, I: exposed item, C: interaction probability, Z: item popularity. 
%Traditional models estimate the correlation $P(C|U, I)$, which confounds the interest matching with popularity bias.
%(a) Causal graph of traditional methods that estimate P(C|U,I); (b) Causal graph that considers item popularity; (c) 
We identify $Z$ as the confounder between $I$ and $C$, and propose deconfounded training with $P(C| do(U, I))$ as the interest matching.}
\vspace{-15pt}
\label{fig:rec_graph}
\end{figure}

To understand how item popularity affects the recommendation process, we resort to the language of \textit{causal graph}~\cite{pearl2009causality} for a qualitative analysis first. Figure \ref{fig:rec_graph}(a) illustrates that traditional methods mainly perform user-item matching to predict the affinity score: $U$ (user node) and $I$ (item node) are the causes, and $C$ is the effect node to denote the interaction probability. An example is the prevalent latent factor model~\cite{mf,lightgcn}, which forms the prediction as the inner product between user embedding and item embedding. Since how a model forms the prediction implies how it assumes the labeled data be generated, this causal graph could also interpret the assumed generation process of the observed interaction data. Item popularity, although exerts a significant influence on data generation process, is not explicitly considered by such coarse-grained modeling methods. 

We next consider how item popularity affects the process, enriching the causal graph to Figure~\ref{fig:rec_graph}(b). Let node $Z$ denote item popularity, which has two edges pointing to $C$ and $I$, respectively. First, $Z\rightarrow C$ means the item popularity directly affects the interaction probability, since many users have herd mentality (\textit{a.k.a.,} conformity), thus tend to follow the majority to consume popular items~\cite{conformity,DICE}. Second, $Z\rightarrow I$ means item popularity affects whether the item is exposed, since recommender system usually inherits the bias in the data and exposes popular items more frequently\footnote{Note that we assume users make interaction choice on the exposed items only, and do not consider other information seeking choices like search which is not the focus of this work. Thus the observed interactions are conditioned on the exposure of the interacted items before. }. Remarkably, we find $Z$ is the common cause for $I$ and $C$, acting as the \textit{confounder}~\cite{pearl2009causality} between the exposed items and the observed interactions.
It means that, item popularity $Z$ affects the observed interaction data with two causal paths: 1) $Z\rightarrow C$ and 2) $Z\rightarrow I \rightarrow C$, wherein the second path contains the bad effect of  bias amplification, since it increases observed interactions of popular items even though they may not well match user interest.  

To remove the bad effect of popularity bias on model training, we need to intervene recommended items $I$ to make them immune to their popularity $Z$. Experimentally, it means we need to alter the exposure policy to make it free from the impact of item popularity and then recollect data, which is however costly and impossible to achieve for academia researchers. Thanks to the progress on causal science, we can achieve the same result with \textit{do-calculus}~\cite{pearl2009causality} without performing interventional experiments. In short, we estimate the user-item matching as $P(C| do(U, I))$ that cuts off the path $Z\rightarrow I$ during training, differing from the correlation $P(C | U, I)$ estimated by existing recommender models that confounds user interest with popularity bias. Through such deconfounded training, $P(C| do(U, I))$ estimates user interest matching on items more accurately than $P(C | U, I)$, removing the spurious correlations between $I$ and $C$ due to the confounder of $Z$.
During the inference stage, we infer the ranking score as $P(C| do(U, I), do(Z))$, intervening the item popularity $Z$ with our desired bias (\eg the forecasted popularity in the testing stage).

The main contributions of this work are summarized as follows:
\begin{itemize}[leftmargin=*]
    %\item Popularity is treated as a confounder, and bad effect of popularity bias is attributed to the existence of the confounder.
    \item We analyze how popularity bias affects recommender system with causal graph, a powerful tool but is seldom used in the community; we identify that the bad impact of popularity bias stems from the confounding effect of item popularity. 
    %build the causal graph of recommendation and identify the popularity as a confounder, which is the root reason of popularity bias's bad effect.
    %We rebuild the causal graph of recommendation generation process, in which popularity is  treated as a confounder. We attribute the bad impact of popularity bias to the existing of the confounder.
    %\item We try to leverage popularity bias instead of blindly pursing unbiased status based on causal intervention. 
    \item We propose \textit{Popularity-bias Deconfounding and Adjusting} (PDA), a novel framework that performs deconfounded training with \textit{do}-calculus and causally intervenes the popularity bias during recommendation inference.  
    %PD and PDA, forming a new paradigm of building recommender system, which is empowered by the causal inference techniques: backdoor adjustment and intervention.
    %We utilize popularity bias from, there are two key aspects: (1) A learning paradigm for recommendation to remove the bad impact of popularity bias based on post-intervention by Backdoor Criterion. (2) Inject one desired popularity bias by intervening the popularity to the desired one..
    %\item We conduct experiments on two public datasets Kwai and Douban and a private dataset Tencent. Extensive  results  demonstrate the power of our methods.
    \item We conduct extensive experiments on three real datasets, validating the effectiveness of our causal analyses and methods.% We will release codes to facilitate future research. 
\end{itemize}

%%% IPS 
%%% causal embedding
%%% reranking or reg
%%% advantage and disadvantages
%%% Both aim to eliminate the pop bias or force to improve dive

%% Our new perspective: pop is confounder. We should not only eliminate pop bias, but also deconfound it leverage it well. 

%% 4. Our method, how we achieve it. Important insights

%% 5. Summary of contributions.

% \begin{figure}
% \centering
% \subfigure[\textbf{Causal graph of traditional methods. $P(C|U,I)$}]{\label{fig:a}\includegraphics[width=0.15\textwidth]{figures/tradition_graph.pdf}}
% \subfigure[ \textbf{Our proposed intervention model estimates $P(C|do(U,I))$}]{\label{fig:b} \includegraphics[width=0.3\textwidth]{figures/our_graph.pdf}}
% \caption{Causal graphs to describe the recommendation process. U: user, I: exposed item, C: interaction probability, Z: item popularity. 
% %Traditional models estimate the correlation $P(C|U, I)$, which confounds the interest matching with popularity bias. 
% We identify $Z$ as the confounder between $I$ and $C$, and propose deconfounded training with $P(C| do(U, I))$ as the interest matching.}
% \label{fig:rec_graph}
% \end{figure}

\section{Primary knowledge}
% In this section, we would like to introduce the symbols that we used for recommendation. And then show that item popularity trends are drift with time goes to show that it's desirable to introduce some bias in recommendation.
We use uppercase character (\eg $U$) to denote a random variable and lowercase character (\eg $u$) to denote its specific value. We use characters in calligraphic font (\eg $\mathcal{U}$) to represent the sample space of the corresponding random variable, and use $P(\cdot)$ to represent probability distribution of a random variable. 

% \paragraph{Problem formulation.} 
Let $\mathcal{D}$ denote the historical data, which is sequentially collected through $T$ stages, \ie $\mathcal{D} = \{\mathcal{D}_1 \cup \cdots \cup \mathcal{D}_T\}$; $\mathcal{U}=\{u_1 , \dots, u_{|\mathcal{U}|}\}$ and $\mathcal{I}=\{i_1 , \dots, i_{{|\mathcal{I}|}}\}$ denote all users and items, respectively. 
%The target of recommender system is to learn the user preference over items from historical user behaviors. 
Through learning on historical data, the recommender system is expected to capture user preference and serves well for the next stage $T+1$. That is, it aims to obtain high recommendation accuracy on $\mathcal{D}_{T+1}$. 
%That is, estimating the probability that a user clicks a item in the future stage $T+1$. 
%In our setting, we investigate the effect of item popularity.% which is a term defined over a set of interactions. 
In this work, we are interested in the factor of item popularity, defining the \textit{local popularity} of item $i$ on the stage $t$ as: 
%Formally, for each stage t, we defined the popularity of item $i$ over $\mathcal{D_t}$ as
%  \begin{equation}\label{eq:pop_define}\small
%      m_{i}^{t} = \frac{D_{i}^{t}}{\sum_{j\in \mathcal{I}} D_{j}^{t}},
%  \end{equation}
 \begin{equation}\label{eq:pop_define}\small
     m_{i}^{t} = D_{i}^{t} / \sum_{j\in \mathcal{I}} D_{j}^{t},
 \end{equation}
where $D_{i}^{t}$ denotes the number of observed interactions for item $i$ in $\mathcal{D}_{t}$. We can similarly define the global popularity of an item $m_i$ based on its interaction frequency in $\mathcal{D}$, but we think the local popularity has a larger impact on the system's exposure mechanism and user decision, since the system is usually periodically retrained and the most recent data has the largest impact. 

\paragraph{Popularity drift.} 
Intuitively, item popularity is dynamic and changes over time, meaning that the impact of popularity bias could also be dynamic. To quantify the popularity drift, 
%Considering that item popularity drifts over time~\cite{MostRecent}, it is necessary to quantify the change of item popularity (\ie popularity drift) to facilitate the study of popularity bias. 
%To achieve the target, 
we define a metric named \textit{Drift of Popularity} (DP) to measure the drift between two stages.
First, we represent each stage $t$ as a probability distribution over items: $\big[m_{1}^{t},\dots,m_{|\mathcal{I}|}^{t}\big]$, where each entry denotes an item's frequency in the stage. 
Then, we use the Jensen-Shannon Divergence (JSD)~\cite{js-range} to measure the similarity between two stages: 
%Considering that item popularity can be treated as a probability distribution over items, we define DP based on Jensen-Shannon Divergence (JSD)~\cite{js-range}, which is widely used to measure the similarity between two probability distributions. Formally:
\begin{equation}\small
    \begin{aligned}
    DP(t,s) = JSD\left(
        \big[m_{1}^{t},\dots,m_{|\mathcal{I}|}^{t}\big],
        \big[m_{1}^{s},\dots,m_{|\mathcal{I}|}^{s}\big]
    \right),
\end{aligned}
\end{equation}
% \begin{align}
%     DP(t,s) = JSD\left(
%         \big[m_{1}^{t},\dots,m_{|\mathcal{I}|}^{t}\big],
%         \big[m_{1}^{s},\dots,m_{|\mathcal{I}|}^{s}\big]
%     \right),
% \end{align}
where $t$ and $s$ are the two stages.
%; $m_i^{t}$ and $m_i^{s}$ represent the corresponding popularity of item $i$. 
Similar to JSD, the range of DP is $[0,log(2)]$, and a higher value indicates a larger popularity drift. 
%Item popularity will drift with time goes by, from the data perspective, popular items will be different in the different stage. To better predict the user's interaction, the popularity drift can be introduced in prediction. The popularity that will drift into is the desired popularity. To quantify the popularity drift between two stages, we define a metric named \textit{Drift of Popularity}(DP) based on Jensen-Shannon Divergence(JSD)~\cite{js-range}. JSD is used to measure the similarity between two probability distributions, and its value drops in $[0,log(2)]$~\cite{js-range}, a higher value means the more differences between the two distribution, and $JSD=0$ means the two distribution are the same. Then the DP can be defined as:
% $$DP(k,l) = JSD([m_{1}^{k},\dots,m_{|\mathcal{I}|}^{k}], [m_{1}^{l},\dots,m_{|\mathcal{I}|}^{l}])$$
% where $k,l$ denote stage $k$ and $l$, $m_i^{k}$ represent the item popularity of i in stage $k$, and $[m_{1}^{k},\dots,m_{|\mathcal{I}|}^{k}]$ can be treated a probability distributions regarding item popularity. 
%And we take Laplace Smoothing~\cite{Laplace-smothing} when computing item popularity to smooth $m$ to avoid zeros appearing. If the item popularity doesn't drift in two stage, the DP will be 0, otherwise, will be great than 0.
% Since JSD cannot process inputs with zeros, Laplace Smoothing~\cite{Laplace-smothing} is applied to item popularity in the calculation of DP.

\begin{figure}
    \centering
    \subfigure[$DP(t,t+1)$]{\includegraphics[width=0.23\textwidth]{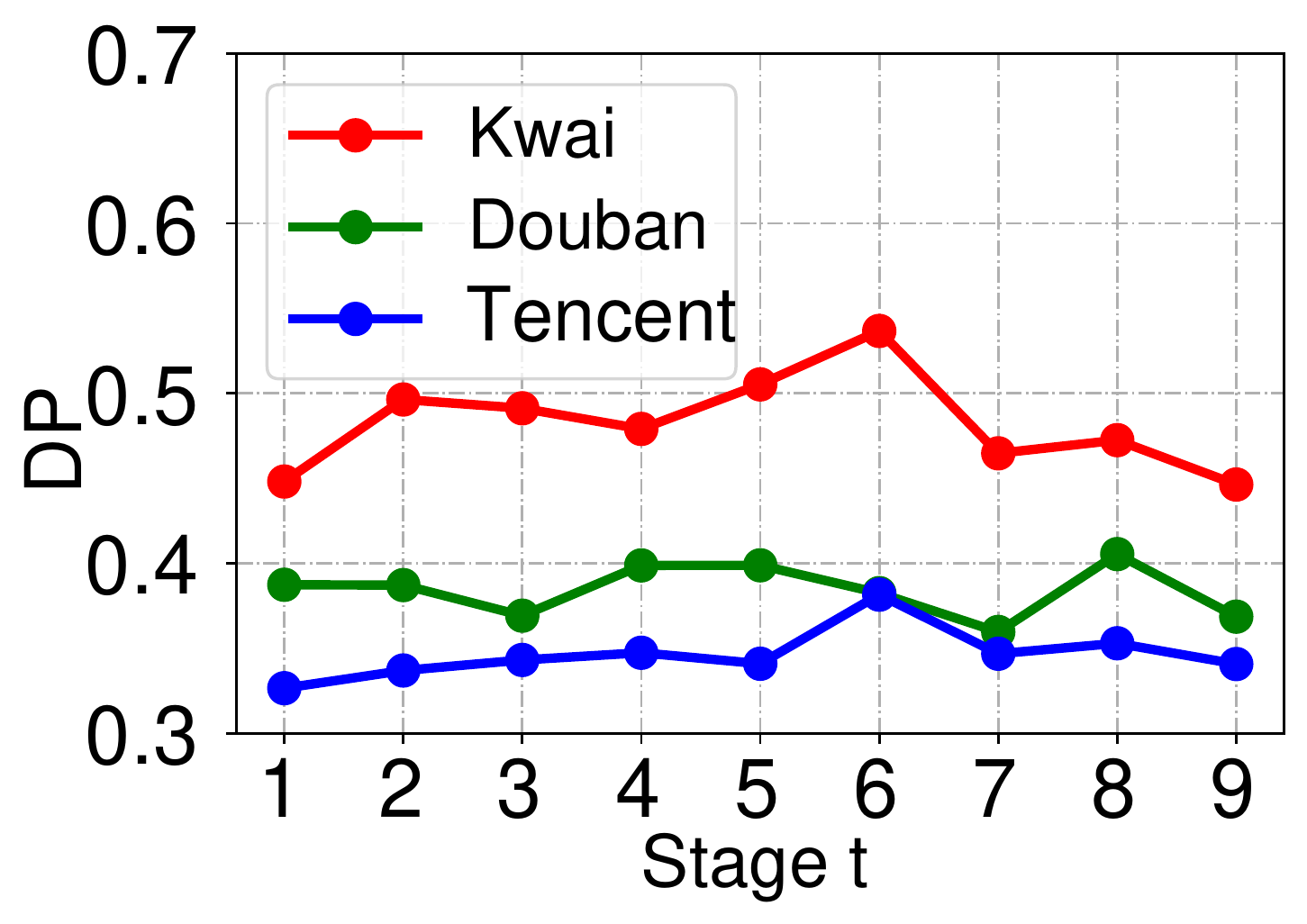}}
    \subfigure[$DP(1,t)$]{\includegraphics[width=0.23\textwidth]{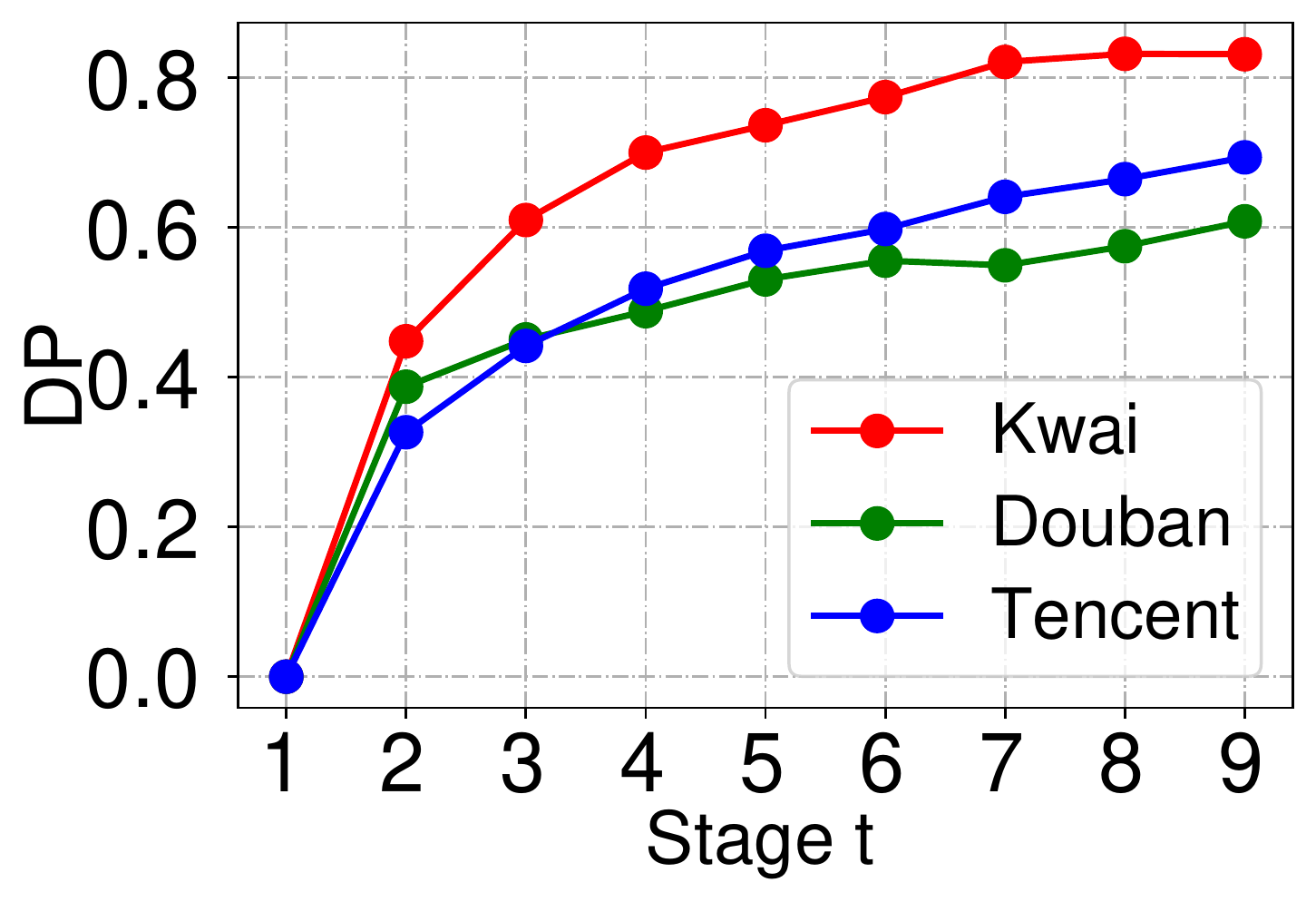}}
    \vspace{-15pt}
    %\caption{Item popularity distribution changes over time measured by JS divergence between different stages: (a) between t-th and $(t+1)$-th period, (b) between 1-th and t-th period.}
    \caption{Popularity drift between: (a) two successive stages $DP(t, t+1)$; (b) the first and present stage  $DP(1, t)$.}
    \vspace{-13pt}
    \label{fig:drit_showing}
\end{figure}

Figure~\ref{fig:drit_showing}(a) shows the DP value of two successive stages, \ie $DP(t,t+1)$ where $t$ iterates from 1 to 9, on three real-world datasets (details in Section~\ref{exp:setting}). We can see that the popularity drift obviously exists in all three datasets, and different datasets exhibit different levels of popularity drift. 
Figure~\ref{fig:drit_showing}(b) shows the DP value of the first and present stage, \ie $DP(1,t)$, which measures the accumulated popularity drift. 
We can see a clear increasing trend, indicating that the longer the time interval is, the large popularity drift the data exhibits. 
%We then conduct a quantitative analysis of the popularity drift in the historical data. In particular, we calculate $DP(t,t+1)$, which reflects the popularity drift between adjacent stages; and $DP(1,t)$, which represents the popularity drift as compared to the first stage. 
%For explicitly showing the popularity drift, we split datasets to 10 stages, then we compute $DV(t,t+1)$ and $DV(1,t)$, where $t={1,\dots,9}$ represent different stages, to show the differences. 
%Figure~\ref{fig:drit_showing} shows the DP measure on three representative recommendation datasets (see Section~\ref{sec:dataset} for details). (1) From Figure~\ref{fig:drit_showing}(a), we can find that the popularity drift obviously exists in all the datasets, and different datasets have different levels of popularity drift. (2) From Figure~\ref{fig:drit_showing}(b), we can see that stages with longer interval exhibit larger popularity drift.
%
These results reveal that popularity bias and its impact also change with time. The popularity bias in the future stage differs from that of past stages. If we set the target of model generalization as pursuing high accuracy on the next stage data $\mathcal{D}_{T+1}$\footnote{To our knowledge, many industrial practitioners on recommender systems use the next stage data for model evaluation.}, then a viable way is to predict popularity trend and inject it into the recommendations.  

\section{Methodology}
%In this section, we first analyze introduce the causal view of interaction generation process (Section~\ref{ssec:causal graph}), then elaborate the proposed PDA framework (Section~\ref{ssec:PD} and~\ref{ssec:PDA}) followed by an in-depth interpretation of the proposed PDA, compared with correlation based methods.
% In this section, we first analyze the reason of popularity bias amplification from a causal view (Section~\ref{ssec:causal graph}), then elaborate the proposed PDA framework (Section~\ref{ssec:PD} and~\ref{ssec:PDA}) followed by an in-depth interpretation of the proposed PDA.
We first analyze the impact of item popularity on recommendation from a fundamental view of causality. Then we present our PDA framework that eliminates the bad effect of popularity bias on training and intervenes the inference stage with desired bias. Lastly, we present in-depth analyses of our method.

\subsection{Causal View of Recommendation}\label{ssec:causal graph}
Figure \ref{fig:rec_graph}(b) shows the causal graph that accounts for item popularity in affecting recommendation. By definition, 
% \subsubsection{Causal graph}
%Causal model is used to descibe the logics of how the data is generated, in which causal graph is used to describe the causal relations between variables. 
%A causal graph can be described as a directed acyclic graph $G = \{V,E\}$, where $V$ is a set of nodes and $E$ is a set of directed edges. 
causal graph~\cite{pearl2009causality} is a directed acyclic graph where a node denotes a variable and an edge denotes a causal relation between two nodes~\cite{pearl2009causality}.
It is widely used to describe the process of data generation, which can guide the design of predictive models. Next, we explain the rationality of this causal graph.

\begin{itemize}[leftmargin=*]
    \item Node $U$ represents the user node, \eg user profile or history feature that is used for representing a user. 
    \item Node $I$ represents the item node. We assume a user can only interact with (\eg click) the items that have been exposed to him/her, thus $I$ denotes the exposed item. 
    \item Node $C$ represents the interaction label, indicating whether the user has chosen/consumed the exposed item.
    \item Node $Z$ represents the item popularity, which can be seen as a hidden variable since it is usually not explicitly modeled by traditional methods (\eg \cite{mf,lightgcn} form predictions with Figure \ref{fig:rec_graph}(a)). But it has a large impact on the recommendation. 
    \item Edges $\{U, I, Z\} \rightarrow C$ denote that an interaction label $C$ is determined by the three factors: user $U$, item $I$, and the item's popularity $Z$. Traditional methods only consider $\{U, I\}\rightarrow C$ which is easy to explain: the matching between user interest and item property determines whether a user consumes an item. Here we intentionally add a cause node $Z$ for capturing user conformity --- many users have the herd mentality and tend to follow the majority to consume popular items~\cite{conformity,DICE}. Thus, whether there is an interaction is the combined effect of $U, I$ and $Z$. 
    \item Edge $Z\rightarrow I$ denotes that item popularity affects the exposure of items. This phenomenon has been verified on many recommender models, which are shown to favor popular items after training on biased data~\cite{Himan-thesis}. 
    %Popular items are more likely to be exposed to users, since they take more proportion in the training data an
\end{itemize}

From this causal graph, we find that item popularity $Z$ is a confounder that affects both exposed items $I$ and observed interactions $C$. This results in two causal paths starting from $Z$ that affect the observed interactions: $Z\rightarrow C$ and $Z\rightarrow I\rightarrow C$. The first path is combined with user-item matching to capture the user conformity effect, which is as expected. In contrast, the second path means that item popularity increases the exposure likelihood of popular items, making the observed interactions consist of popular items more. Such effect causes bias amplification, which should be avoided since an authentic recommender model should estimate user preference reliably and is immune to the exposure mechanism.

In the above text, we have explained the causal graph from the view of data generation. In fact, it also makes sense to interpret it from the view of model prediction. Here, we denote $U$ and $I$ as user embedding and item embedding, respectively, and traditional models apply inner product~\cite{mf,lightgcn} or neural network~\cite{ncf} above them for prediction (\ie Figure \ref{fig:rec_graph}(a)). Item popularity $Z$ is not explicitly considered in most models, however, it indeed affects the learning of item embedding. For example, it increases the vector length of popular items, making inner product models score popular items high for every user~\cite{DICE}. Such effect justifies the edge $Z\rightarrow I$, which exerts a negative impact to learn real user interest. 

To sum up, regardless of which explanation for the causal graph, $Z\rightarrow I$ causes the bad effect and should be eliminated in formulating the predictive model. 

\subsection{Deconfounded Training}\label{ssec:PD}
We now consider how to obtain a model that is immune to the impact of $Z\rightarrow I$. Intuitively, if we can intervene the exposure mechanism to make it randomly expose items to users, then the collected interaction data is free from the impact of $Z\rightarrow I$. Directly training traditional models on it will do. However, the feasibility and efficacy of such solutions are low: first, only the recommender builder can intervene the exposure mechanism, and anyone else (\eg academia researcher) has no permission to do it; second, even for recommender builder that can intervene the exposure mechanism, they can only use small random traffic, since random exposure hurts user experience much. Effective usage of small uniform data remains an open problem in recommendation research~\cite{2018causalE, distillation}. 

Fortunately, the progress of causal science provides us a tool to achieve intervention without performing the interventional experiments~\cite{pearl2009causality}. The secret is the \textit{do}-calculus. In our context, performing $do(I)$ forces to remove the impact of $I$'s parent nodes, achieving our target. As such, we formulate the predictive model as $P(C| do(U, I) )$, rather than $P(C| U, I )$ estimated by traditional methods\footnote{Note that since there is no backdoor path between $U$ and $C$ in our causal graph, it holds that $P(C| U, do(I) ) = P(C| do(U), do(I) ) = P( C| do (U, I))$. Here we use $P(C| do(U, I) )$ as the intervention model for brevity. }. %\footnote{Note that since $U$ has no parents in our causal graph, it holds that $P(C| U, do(I) ) = P(C| do(U), do(I) ) = P( C| do (U, I))$. Here we use $P(C| do(U, I) )$ as the intervention model for brevity. }. 

Let the causal graph shown in Figure~\ref{fig:rec_graph}(b) be $G$, and the intervened causal graph shown in Figure~\ref{fig:rec_graph}(c) be $G'$. Then, performing \textit{do}-calculus on $G$ leads to:
\begin{equation}\label{eq:do-derivation}\small
	\begin{split}
	P(C|do(U,I)) & \overset{(1)}{=}  P_{G'} (C|U,I)\\
	& \overset{(2)}{=} \sum_z P_{G'}(C|U,I,z)P_{G'}(z|U,I)\\
	& \overset{(3)}{=} \sum_z P_{G'}(C|U,I,z)P_{G'}(z)\\
	& \overset{(4)}{=} \sum_z P(C|U,I,z) P(z),
	\end{split}
\end{equation}
where $P_{G'}(\cdot)$ denotes the probability function evaluated on $G'$. Below explains this derivation step by step: 
\begin{itemize}[leftmargin=*]
    \item (1) is because of \textit{backdoor criterion}~\cite{pearl2009causality} as the only backdoor path $I\leftarrow Z \rightarrow C$ in $G$ has been blocked by $do(U, I)$;
    \item (2) is because of Bayes' theorem;
    \item (3) is because that $U$ and $I$ are independent with $Z$ in $G'$; 
    \item (4) $P(C|U,I,Z) = P_{G'}(C|U,I,Z)$ is because that the causal mechanism $\{U,I,Z\}\rightarrow C$ is not changed when cutting off $Z\rightarrow I$, $P(Z) = P_{G'}(Z)$ since $Z$ has the same prior on the two graphs.
\end{itemize}

Next, we consider how to estimate $P(C|do(U, I))$ from data. Apparently, we need to first estimate $P(C|U,I,Z)$, and then estimate $\sum_z P(C|U,I,z) P(z)$. \vspace{+5pt}

\noindent \textbf{Step 1. Estimating $P(C|U,I,Z)$.} This conditional probability function evaluates that given a user-item pair $U=u, I=i$ and the item's present popularity as $Z=m_i^t$, how likely the user will consume the item. Let the parameters of the conditional probability function be $\Theta$, we can follow traditional recommendation training to learn $\Theta$, for example, optimizing the pairwise BPR objective function on historical data $\mathcal{D}$: 
\begin{equation}\label{eq:bpr}\small
    \max_{\Theta} \sum_{(u, i, j) \in \mathcal{D}} \text{log} \sigma \left(
        P_\Theta\left(
            c=1 | u, i, m_{i}^{t}
        \right) - P_\Theta \left(
            c=1 | u, j, m_{j}^{t}
        \right) 
    \right),
\end{equation} 
where $j$ denotes the negative sample for $u$, $\sigma(\cdot)$ is the sigmoid function. The $L_2$ regularization is used but not shown for brevity.  

Having established the learning framework, we next consider how to parameterize $P_\Theta(c=1 | u, i, m_{i}^{t})$. Of course, one can design it as any differentiable model, factorization machines, or neural networks. But here our major consideration is to decouple the user-item matching with item popularity. The benefits are twofold: 1) decoupling makes our framework extendable to any collaborative filtering model that concerns user-item matching; and 2) decoupling enables fast adjusting of popularity bias in the inference stage (see Section~\ref{sec:PDA}), since we do not need to re-evaluate the whole model. To this end, we design it as:
\begin{equation} \label{eq:model}\small
    P_\Theta(c=1 | u, i, m_{i}^{t}) = ELU'(f_\Theta(u, i)) \times ({m}_{i}^{t})^\gamma,
\end{equation}
where $f_\Theta(u, i)$ denotes any user-item matching model and we choose the simple \textit{Matrix Factorization} (MF) in this work; hyper-parameter $\gamma$ is to smooth the item popularity and can control the strength of conformity effect: setting $\gamma=0$ means no impact and larger value assigns larger impact. $ELU'(\cdot)$ is a variant of the Exponential Linear Unit \cite{elu} activation function that ensures the positivity of the matching score:
\begin{equation} \label{eq:alph-example}\small
	ELU'(x) = \begin{cases}
	e^{x}, &\text{if } x \leq 0 \\ 
	x+1, &\text{else}  %\in \mathbb{R}\setminus\mathbb{Q}	
		   \end{cases}
\end{equation}
This is to ensure the monotonicity of the probability function since $({m}_{i}^{t})^\gamma$ is always a positive number. Lastly, note that one needs to normalize $ P_\Theta(c=1 | u, i, m_{i}^{t})$ to make it a rigorous probability function, but we omit it since it is time-consuming and does not affect the ranking of items. \vspace{+5pt}

% \begin{algorithm}[t]
% 	\caption{Popularity-bias Deconfounding(PD)}
% 	\LinesNumbered
% 	\label{alg:PD-learing}
% 	\KwIn{Dataset $\mathcal{D}=\{D_1,\dots,D_{T}\}$}
% % 	\KwOut{Updated recommender $W_{t+1}$}
% % 	Split D to T chunks \{$D_1,\dots,D_{T-1}$\} along time \;
%     // Step 1: estimate $P(C|U,I,Z)$ \;
% 	Compute each item $i$ popularity $m_{i}^{t}$ on each stage $t$ by Equation \eqref{eq:pop_define} \;
% 	Match each interaction $(u,i)$ in stage $t$ with $m_{i}^{t}$\;
% 	\While{Stop condition is not reached}{
% 	Randomly sample one batch interactions $B$ in $\mathcal{D}$\;
% 	Compute $f(u,i,m_{i}^{t})$\;
% 	Update model parameters by optimizing Equation~\ref{eq:loss}. 
% 	}
% 	// Step 2: inference with $P(C|do(U,I))$ \;
% 	Ranking items with $\alpha(k_{u,i})$
	
% \end{algorithm}

\noindent \textbf{Step 2. Estimating  $\sum_z P(C|U,I,z) P(z)$.} Now we move forward to estimate the interventional probability $P(C|do(U, I))$. Since the space of $Z$ is large, it is inappropriate to sum over its space for each prediction evaluation. Fortunately,  we can perform the following reduction to get rid of the sum:
%With the estimation of $P(C|U,I,Z)$, we can approximate this intervention probability as:
\begin{equation}\label{eq:doUI}\small
   \begin{split}
    P(C|do(U,I)) &= \sum_{z} P(C|U,I,z)P(z) \notag\\
    & = \sum_z ELU'(f_{\Theta}(u,i)) \times z^{\gamma} P(z) \\
    & = ELU'(f_{\Theta}(u,i)) \sum_z z^{\gamma}P(z) \\
    & = ELU'(f_{\Theta}(u,i)) E(Z^{\gamma})
\end{split} 
\end{equation}
where $E(Z^{\gamma})$ denotes the expectation of $Z^{\gamma}$. Note that the expectation of one variable is a constant. Since $P(C|do(U, I))$ is used to rank items for a user, the existence of $E(Z^{\gamma})$ does not change the ranking. We can thus use $ELU'(f_{\Theta}(u,i))$ to estimate $P(C|do(U, I)$. \vspace{+5pt}

%the goal of recommendation is to rank items for a user. Ranking with $P(C|do(U,I))$ is equal to ranking with $ELU(f_{\Theta}(u,i))$.

\noindent To summarize, we fit the historical interaction data with $P_\Theta(c=1|u, i, m_i^t)$, and use the user-item matching component $ELU'(f_{\Theta}(u,i))$ for deconfounded ranking. We name this method as \textit{Popularity-bias Deconfounding} (PD).

% Considering that we are unable to enumerate the instance of $Z$, we first derive an approximation of $P(C|do(U,I))$. Formally,
% \begin{align}
%     P(C|do(U,I)) &= \sum_z P(C|U,I,z)P(z) \notag\\
%     & \approx P(C|U,I,E(Z)),
% \end{align}
% where $E(Z)$ denotes the expectation of $Z$. It should be noted that $E(Z)$ is constant and not depends on the value of $I$. We can thus approximate $P(C|U,I,E(Z))$ by setting the $m_i^t$ in $h(u,i,m_i^t)$ with a proper constant value, \ie $h(u,i,m^*)$. As the value of $m^*$ does not change the ranking of items, we simply set $m^* = 1$. That is to say, we estimate $P(C|do(U,I))$ as $\alpha(k_{ui})$.

%%%
% TODO: fill in the contents and rephrase
%%%
% In Algorithm \ref{alg:PD-learing}, we summarize the process of PD, which has two main differences to the existing methods:
% \begin{itemize}[leftmargin=*]
%     \item PD estimates $P(C|U,I,Z)$ during recommender training instead of $P(C|U,I)$.
%     \item PD makes recommendation with $P(C|do(U,I))$ instead of $P(C|U,I)$.
% \end{itemize}

%At last, we summarize the process of our PD in Algorithm \ref{alg:PD-learing}.
%  \subsection{Inference with drifted popularity}

%%%
% TODO: update the contents to be coherent with the previous subsection
%%%
\subsection{Adjusting Popularity Bias in Inference}\label{sec:PDA}
Owing to $P(C|do(U, I))$, we can eliminate the bad effect of popularity bias. We now seek for better usage of popularity bias, \eg promoting the items that have the potential to be popular. 
Assume the target popularity bias is $\Tilde{z}$, and we want to endow the recommendation policy with this bias. To achieve the target, we just need to do the intervention of $Z=\Tilde{z}$ for model inference:
\begin{align}\small\label{eq:do-z}
    P(C|do(U=u,I=i),do(Z=\Tilde{z})) = P_\Theta(c=1|u,i, \Tilde{m}_i),
\end{align}
where $\Tilde{m}_i$ denotes the popularity value of $\Tilde{z}$. 
This intervention probability directly equals the conditional probability since there is no backdoor path between $Z$ and $C$ in the causal graph. %$Z$ has no parents in the causal graph.
%Since the derivation step is similar to the Equation~\eqref{eq:do-derivation}, we omit it here. According to the formulation of Step 1, we can approximate it with $P_{\Theta}(C|u,i, \Tilde{z})$. Assume the desired popularity for item $i$ is $\Tilde{m_i}$,  $P_{\Theta}(u,i,\Tilde{m}_i)$ is used for inference. 
% Considering the need of introducing desired popularity bias into the system, such as promoting the items that have the potential to be popular, the core is equal to adjust the popularity be the same to the desired one by force. To realize this goal, we can intervene $Z=\Tilde{z}$ that corresponds to the desired item popularity, and can intervene $U=u$ and $I=i$ to simulate the responses,\ie $P(C|do(U=u,I=i),do(Z=\Tilde{z}))$. Then we can recommend item with the responses to do real recommendation to get higher recommendation accuracy. For  $P(C|do(U=u,I=i),do(Z=\Tilde{z}) )$, according to back-door adjustment, we have:  %we only need to cut down the paths that arrow into $Z$ of the right graph in figure~\ref{fig:rec_causal_pre_now}.There are not any these types of paths. So we can compute it still on the right graph, we can get:
%  \begin{equation}
%  P(C|do(U=u,I=i),do(Z= \Tilde{z} )) = P(C|u,i,\Tilde{z} ) 
%  \end{equation}
Since the focus of this work is not on popularity prediction, we employ a simple time series forecasting method to set $\Tilde{m}_i$: 
\begin{equation}\small \label{eq:pre_pop}
    \Tilde{m}_i = m_{i}^{T} + \alpha (m_{i}^{T} - m_{i}^{T-1}),
\end{equation}
where $m_{i}^{T}$ is the popularity value of the last stage, and $\alpha$ is the hyper-parameter to control the strength of popularity drift in predicting the future. We name this method as \textit{Popularity-bias Deconfounding and Adjusting} (PDA). 

\begin{figure}
    \centering
    \includegraphics[width=0.35\textwidth]{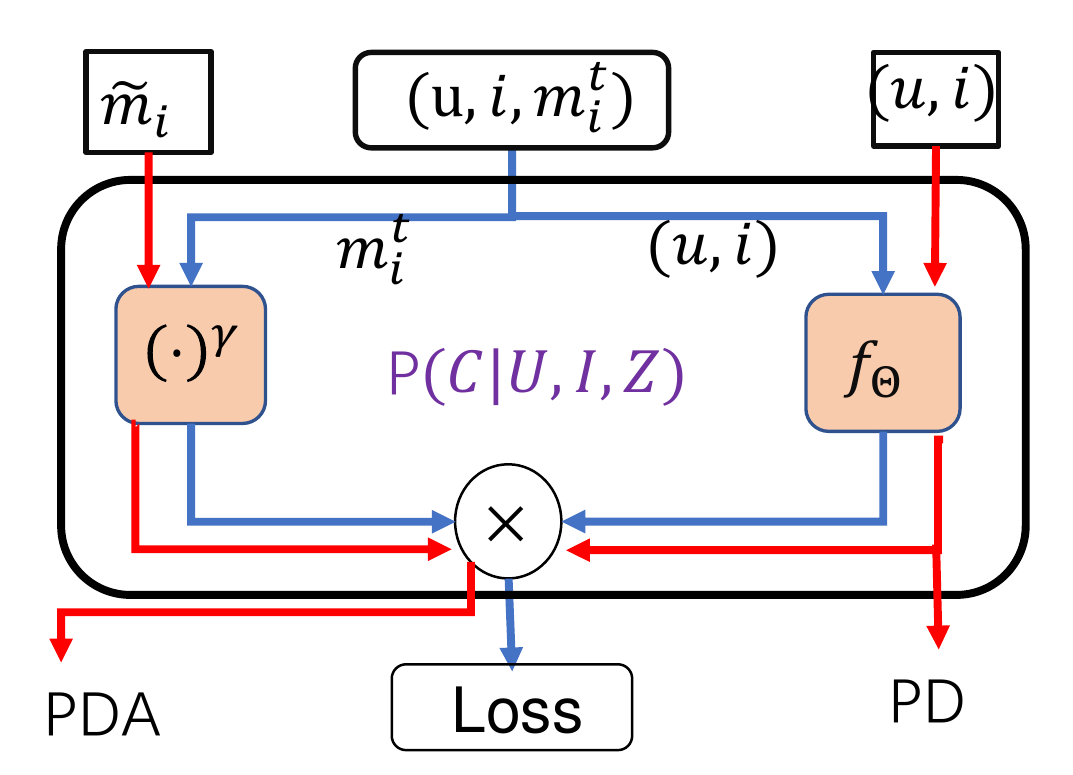} %framwork-nosampling
    \vspace{-0.4cm}
    \caption{The workflow of our method. The blue arrows represent the training stage and the red arrows represent the inference stage.}
    \label{fig:framework}
    \vspace{-13pt}
\end{figure}

%Both Figure \ref{fig:framework} and Algorithm \ref{alg:PD-PDA} summarize our method. The training stage optimizes Equation~(\ref{eq:bpr}) (line 2 in the algorithm). The inference stage can adjust the popularity bias (line 7 in the algorithm) as:
Figure \ref{fig:framework} illustrates the workflow of PDA where the training stage optimizes Equation~(\ref{eq:bpr}) and the inference stage can adjust the popularity bias as:
% \begin{equation}\label{eq:mf}\small
%      PDA_{ui} = ELU'(f_\Theta(u, i)) \times (\Tilde{m}_i)^{\Tilde{\gamma}}.
% \end{equation}
\begin{equation}\label{eq:mf}\small
     PDA_{ui} = ELU'(f_\Theta(u, i)) \times (\Tilde{m}_i)^{\Tilde{\gamma}}.
\end{equation}
$\Tilde{\gamma}$ denotes the popularity smoothing hyper-parameter used for model inference, which can be different from that used in training.
This is to consider that the strength of popularity bias can drift in the future. 
In practice, setting $\Tilde{\gamma} = \gamma$ can achieve expected performance (\cf Table~\ref{tab:PDA-result}). 
Setting $\Tilde{\gamma}=0$  degrades the method to PD that uses only interest matching for recommendation. Considering the importance of properly stopping model training, PDA uses a model selection strategy based on the adjusted recommendation, which is slightly different from that of PD (see Algorithm \ref{alg:PD-PDA}).
% Ideally, if we have the ground truth of $P(C|do(U,I))$ to select a best model, we think that the model parameters $\Theta$ trained for PD can be used for PDA directly. But there is not the ground-truth. In practice, we will tune the hyper-parameter $\gamma$ in training again and select a more appropriate $\Theta$ for PDA (line 4 of the algorithm). Meanwhile, for simplicity, we keep $\Tilde{\gamma}$ as the newfound one.

%Note that we do not use more complicated time-series forecasting methods such as Recurrent Neural Network~\cite{}, since we find that the simple algorithm is sufficient to achieve good results (\cf Table/Figure XX).
%%%
% TODO: need to explain why don't use more complex models.
%%%
%%%
% TODO: view (a) as an ablation study without the forecast
%%%
%We take two methods to predict the desired popularity for $\mathcal{D}_{T+1}$: (a) the first is that treat the item popularity in $D_{t}$ as the predicted popularity, With the fact that the item popularity in the most near stage has the small popularity drift as showing in~\ref{fig:drit_showing}. (b) the second is $\Tilde{m}_i = m_{i}^{T} + \alpha (m_{i}^{T} - m_{i}^{T}) $, in this form, besides considering the most recent item popularity as a import information, taking the item popularity difference on the most recent two stage term as drifting trends.   

%%%
% TODO: add framework figure here
%%%

\subsection{Comparison with Correlation $P(C|U, I)$}
In the beginning, we argue that traditional methods estimate the correlation $P(C|U, I)$ and suffer from the bad effect of popularity bias amplification. Here we analyze the difference between $P(C|U, I)$ and $P(C|do(U, I))$ to offer more insights on this point.

We first transform $P(C|U, I)$ with the following steps:
\begin{equation}\small
    \begin{split}
        P(C|U,I) & \overset{(1)}{=} \sum_{z} P(C,z|U,I) \\
        & \overset{(2)}{=} \sum_{z} P(C|U,I,z)P(z|U,I) \\ 
        & \overset{(3)}{=} \sum_{z} P(C|U,I,z)P(z|I) \\
        & \overset{(4)}{\propto	} \sum_{z} P(C|U,I,z)P(I|z)P(z)  
    \end{split}
\end{equation}
(1) is the definition of marginal distribution; (2) is because of the Bayes' theorem; (3) is because $U$ is independent to $Z$ according to the causal graph; (4) is because of the Bayes' theorem. 

Comparing with $P(C|do(U,I)) = \sum_{z} P(C|U,I,z) P(z)$,
% \begin{equation}\nonumber \small
%   P(C|do(U,I)) = \sum_{z} P(C|U,I,z) P(z). 
% \end{equation}
% $$P(C|do(U,I)) = \sum_{Z} P(C|U,I,Z) P(Z). $$ 
we can see that $P(C|U, I)$ has an additional term: $P(I|Z)$, which fundamentally changes the recommendation scoring. Suppose $I=i$ is a popular item, $P(C|U, I=i, Z=m_i^t)$ is a large value since $m_i^t$ is large, which means the term has a large contribution to the prediction score. $P(I=i | Z=m_i^t)$ is also a large value due to the popularity bias in the exposure mechanism. Multiplying the two terms further enlarges the score of $i$, which gives $i$ a higher score than it deserves. As a result, the popularity bias is improperly amplified. This comparison further justifies the rationality and necessity of $P(C|do(U,I))$ for learning user interests reliably.

\begin{algorithm}[t]
	\caption{PD/PDA}
	\LinesNumbered
	\label{alg:PD-PDA}
	\KwIn{dataset $\mathcal{D}=\{(u,i,m_{i}^{t})\}$; hyper-parameter $\gamma$; predicted popularity $\{\Tilde{m}_i\}$; \textit{mode}: PD or PDA }
	\While{stop condition is not reached}{
	Update model parameters $\Theta$ by optimizing Equation~\eqref{eq:bpr}\;
	\lIf{mode==PD}{validate model with $ELU^{'}(f_\Theta (u,i))$}
	\lElse{validate model with Equation~\eqref{eq:mf} (simplify $\Tilde{\gamma}=\gamma$)}
% 	\If{PD}{validate model with $ELU^{'}(f_\Theta (u,i))$\;}
% 	\Else{validate model with Equation~\eqref{eq:mf}\;}
% 	\ifthenelse{PD}{A}{B}
	}
% 	// Step 2: inference with $P(C|do(U,I))$ \;
    % // testing\;
    \lIf{mode==PD}{recommend items using $ELU^{'}(f_\Theta (u,i))$}
    \lElse{recommend items using Equation~\eqref{eq:mf} (simplify $\Tilde{\gamma}=\gamma$)}
\end{algorithm}

\section{Experiments}
%In this section, we conduct experiments to answer the following questions:
In this section, we conduct experiments to answer two main questions:
% \begin{itemize}
%     \item[\textbf{RQ1:}] How is the performance of our deconfounding method PD, compared with the state-of-the-art methods? Does PD achieve the goal of removing bad effect of popularity bias and keeping good effect of popularity bias?
%     %%%  reducing popularity bias, 应该
%     \item[\textbf{RQ2:}] Can our proposed method PDA inject popularity more effective than other methods?
%     % \item[\textbf{RQ3:}] How does the CNN architecture affect the transfer network?
%     % \item[\textbf{RQ4:}] Where are the improvements of SML come from?
% \end{itemize}
%
\textbf{RQ1:} Does PD achieve the goal of removing the bad effect of popularity bias?
How is its performance compared with existing methods? 
    %%%  reducing popularity bias, 应该
\textbf{RQ2:} Can PDA effectively inject desired popularity bias? To what extent leveraging popularity bias enhance the recommendation performance.
    % \item[\textbf{RQ3:}] How does the CNN architecture affect the transfer network?
    % \item[\textbf{RQ4:}] Where are the improvements of SML come from?

%We present experimental settings, followed by results and analyses to answer the above research questions.

\begin{table*}[]
\caption{Recommendation performance after deconfounded training on the three datasets. ``RI'' refers to the relative
improvement of PD over the corresponding baseline. The best results are highlighted in bold and sub-optimal results are underlined.}
\vspace{-8pt}
\label{tab:PD-result}
\resizebox{0.8\textwidth}{!}{%
\begin{tabular}{cccccccccccc}
\toprule
\multirow{2}{*}{Dataset} & \multicolumn{1}{c|}{\multirow{2}{*}{Methods}} & \multicolumn{5}{c|}{Top 20} & \multicolumn{5}{c}{Top 50} \\
 & \multicolumn{1}{c|}{} & Recall & \multicolumn{1}{l}{Precision} & \multicolumn{1}{l}{HR} & \multicolumn{1}{l}{NDCG} & \multicolumn{1}{l|}{RI} & \multicolumn{1}{l}{Recall} & \multicolumn{1}{l}{Precision} & \multicolumn{1}{l}{HR} & \multicolumn{1}{l}{NDCG} & \multicolumn{1}{l}{RI} \\ \hline
\multicolumn{1}{c|}{\multirow{6}{*}{Kwai}} & MostPop & 0.0014 & 0.0019 & 0.0341 & 0.0030 & 632.4\% & 0.0040 & 0.0021 & 0.0802 & 0.0036 & 480.9\% \\
\multicolumn{1}{c|}{} & BPRMF & 0.0054 & {\ul 0.0057} & 0.0943 & 0.0067 & 146.3\% & 0.0125 & {\ul 0.0053} & 0.1866 & 0.0089 & 121.0\% \\
\multicolumn{1}{c|}{} & xQuad & 0.0054 & {\ul 0.0057} & 0.0948 & 0.0068 & 145.0\% & 0.0125 & {\ul 0.0053} & 0.1867 & 0.0090 & 120.3\% \\
\multicolumn{1}{c|}{} & BPR-PC & {\ul 0.0070} & 0.0056 & {\ul 0.0992} & {\ul 0.0072} & 125.0\% & {\ul 0.0137} & 0.0046 & 0.1813 & {\ul 0.0092} & 123.7\% \\
\multicolumn{1}{c|}{} & DICE & 0.0053 & 0.0056 & 0.0957 & 0.0067 & 147.8\% & 0.0130 & 0.0052 & {\ul 0.1872} & 0.0090 & 119.0\% \\
\multicolumn{1}{c|}{} & PD & \textbf{0.0143} & \textbf{0.0138} & \textbf{0.2018} & \textbf{0.0177} & \textbf{-} & \textbf{0.0293} & \textbf{0.0118} & \textbf{0.3397} & \textbf{0.0218} & - \\ \hline
\multicolumn{1}{c|}{\multirow{6}{*}{Douban}} & MostPop & 0.0218 & 0.0297 & 0.2373 & 0.0349 & 75.4\% & 0.0490 & 0.0256 & 0.3737 & 0.0406 & 55.9\% \\
\multicolumn{1}{c|}{} & BPRMF & 0.0274 & {\ul 0.0336} & 0.2888 & 0.0405 & 47.0\% & 0.0581 & {\ul 0.0291} & 0.4280 & {\ul 0.0475} & 34.3\% \\
\multicolumn{1}{c|}{} & xQuad & 0.0274 & {\ul 0.0336} & {\ul 0.2895} & 0.0391 & 48.3\% & 0.0581 & {\ul 0.0291} & {\ul 0.4281} & 0.0473 & 34.4\% \\
\multicolumn{1}{c|}{} & BPR-PC & {\ul 0.0282} & 0.0307 & 0.2863 & 0.0381 & 51.6\% & {\ul 0.0582} & 0.0271 & 0.4260 & 0.0457 & 38.0\% \\
\multicolumn{1}{c|}{} & DICE & 0.0273 & {\ul 0.0336} & 0.2845 & {\ul 0.0421} & 46.2\% & 0.0513 & 0.0273 & 0.4000 & 0.0460 & 44.5\% \\
\multicolumn{1}{c|}{} & PD & \textbf{0.0453} & \textbf{0.0454} & \textbf{0.3970} & \textbf{0.0607} & \textbf{-} & \textbf{0.0843} & \textbf{0.0362} & \textbf{0.5271} & \textbf{0.0686} & - \\ \hline
\multicolumn{1}{c|}{\multirow{6}{*}{Tencent}} & MostPop & 0.0145 & 0.0043 & 0.0684 & 0.0093 & 340.8\% & 0.0282 & 0.0035 & 0.1181 & 0.0135 & 345.8\% \\
\multicolumn{1}{c|}{} & BPRMF & 0.0553 & {\ul 0.0153} & 0.2005 & 0.0328 & 27.1\% & 0.1130 & {\ul 0.0129} & 0.3303 & 0.0497 & 25.3\% \\
\multicolumn{1}{c|}{} & xQuad & 0.0552 & {\ul 0.0153} & 0.2007 & 0.0326 & 27.3\% & 0.1130 & {\ul 0.0129} & 0.3302 & 0.0497 & 25.3\% \\
\multicolumn{1}{c|}{} & BPR-PC & {\ul 0.0556} & {\ul 0.0153} & {\ul 0.2018} & {\ul 0.0331} & 26.5\% & {\ul 0.1141} & 0.0128 & {\ul 0.3322} & {\ul 0.0500} & 24.9\% \\
\multicolumn{1}{c|}{} & DICE & 0.0516 & 0.0149 & 0.1948 & 0.0312 & 32.8\% & 0.1010 & 0.0132 & 0.3312 & 0.0486 & 29.0\% \\
\multicolumn{1}{c|}{} & PD & \textbf{0.0715} & \textbf{0.0195} & \textbf{0.2421} & \textbf{0.0429} & \textbf{-} & \textbf{0.1436} & \textbf{0.0165} & \textbf{0.3875} & \textbf{0.0641} & - \\ \hline
\end{tabular}%
}
\end{table*}

\subsection{Experimental Settings}
\subsubsection{Datasets}\label{exp:setting}
%  \label{exp:setting}
% \subsubsection{Datasets}\label{sec:dataset}
% \noindent $\bullet$ \textit{Datasets.}

We conduct experiments on three datasets:

1) \textbf{Kwai}: This dataset was adopted in Kuaishou User Interest Modeling Challenge\footnote{https://www.kuaishou.com/activity/uimc}, which contains click/like/follow records between users and videos. It spans about two weeks. In this paper, we only utilize clicking data. Following previous work~\cite{ngcf}, we take 10-core filtering to make sure that each user and each item have at least 10 interacted records. After filtering, the experimented data has 7,658,510 interactions between 37,663 users and 128,879 items.

2) \textbf{Douban Movie}: This dataset is collected from a popular review website Douban\footnote{www.douban.com} in China by~\cite{douban}. It contains user ratings for movies and spans more than ten years. We take all rating records as positive samples and only utilize the data after 2010. The 10-core filtering is also adopted. After filtering, there are 7,174,218 interactions between 47,890 users and 26,047 items.
%
%We collected users' ratings in three domains(i.e., movie, book and music) from Douban(www.douban.com), which is a popular review website in China.

3) \textbf{Tencent}: This dataset is collected from Tencent short-video platform from Dec 27 2020 to Jan 6 2021. In this dataset, user interactions are likes, which are reflective of user satisfaction but far more sparse than clicks. Because of its high sparsity, we adopt the 5-core filtering setting, obtaining 1,816,046 interactions between 80,339 users and 27,070 items.

%For all models, we split the data into same training and testing sets for fair comparison. \citet{FollowCrowd} suggest that even though random data splitting is a very common practice in recommendation studies, splitting data along its temporal signal suits the real scenario better. Hence, we split the data samples according to the timestamp of exposure events along the timeline into multiple time stages. Additionally, we don't take uniform sampling among items~\cite{DICE}, because the overall goal of deconfounding is to reflect users' real interest instead of ensuring each item an equal chance of being recommended. Moreover, it is also impossible in practical recommendation to expose each item an equal number of times.
%
%To test the model performance without the influence of popularity bias of training set, temporal splitting is used. 
Following \citet{FollowCrowd}, we split the datasets chronologically. In particular, we split the datasets into 10 time stages according to the interaction time, and each stage has the same time interval. The last stage is left for validation and testing, in which $30\%$ of users are regarded as validation set and the other $70\%$ of users are regarded as testing set. Note that we do not consider new users and new items in validation and testing.
%we don't take uniform sampling among items~\cite{DICE}, because the overall goal of deconfounding is to reflect users' real interest instead of ensuring each item an equal chance of being recommended.

%Data of different stages are denoted as $\{\mathcal{D}_1 ,\dots,\mathcal{D}_{10} \}$. We set $D_{train} =\{\mathcal{D}_1 ,\dots,\mathcal{D}_9 \}$. The $\mathcal{D}_{10}$ is left for validation and testing, in which $30\%$ of users are regarded as validation set and the other $70\%$ of users are regarded as testing set.

% \noindent $\bullet$ \textit{Baselines.}
\subsubsection{Baselines}
We compare PD with the following baselines:

\textbf{- MostPop.} This method simply recommends the most popular items for all users without considering personalization.

\textbf{- BPRMF,} which optimizes the MF model with BPR loss~\cite{bpr}.

\textbf{- xQuAD~\cite{xquad2019}.} This is a Ranking Adjustment method that aims to increase the coverage of long-tail items in the recommendation list. We use the codes released by the authors, and utilize it to re-rank the result of BPRMF. Following its original setting, we tune the bias controlling hyper-parameter $\lambda$ in $[0,1]$ with step 0.1.

\textbf{- BPR-PC~\cite{BPR_PC}.} This is a state-of-the-art Ranking Adjustment method for controlling popularity bias. It has two choices: re-ranking and regularization. Here we implement the re-ranking method based on the BPRMF, because this method shows better performance in the original paper. Following the paper, we tune the bias controlling hyper-parameters $\alpha$ [0.1, 2.0] with step 0.2, and $\beta$ in [0, 1] with step 0.2.

\textbf{- DICE~\cite{DICE}.} This is a state-of-the-art method for learning causal embeddings to handle the popularity bias. 
%The results show DICE can outperform CauseE~\cite{2018causalE} and IPS~\cite{IPS_rec,rec_ips_cap_norm} methods. 
It designs a framework with causality-oriented data to disentangle user interest and conformity into two sets of embedding. 
%According to the their paper, it can work well without balanced data. 
We use the codes provided by the authors, and we take the settings suggested for large datasets instead of the default settings since our datasets are relatively large.
%, and keep its optimal setting except replacing the regularization term $L_{discrepancy}$ from $dCor$ with another option --- $L2$. Since with our data scale, computing $dCor$ will be out of memory for the 2080Ti GPU. It is also suggested in their paper.
Because DICE demonstrates superior performance over IPS-based methods~\cite{IPS_rec,rec_ips_cap_norm}, we do not include IPS-based methods as baselines.

To evaluate the effect of PDA on leveraging desired popularity bias, we compare it with the following popularity-aware methods:
%%%  现在方法只提到了一种方法,我们需要使用一种predicted,然后另一种直接用D_T的popularity

\textbf{- MostRecent~\cite{MostRecent}}. This is an improved version of MostPop, which recommends the most popular items in the last stage rather than in the entire history. Note that PDA takes the last stage to forecast the popularity of next stage to make adjustment.

\textbf{- BPRMF(t)-pop~\cite{temp_pop2017}}. This method models the temporal popularity by training a set of time-specific parameters for each stage. We directly utilize the parameters corresponding to the last training stage for validation and testing.

\textbf{- BPRMF-A} and \textbf{DICE-A}. We enhance BPRMF and DICE to inject desired popularity bias during inference in the same manner as PDA by substituting $f_{\Theta}(u,i)$ with the prediction of BPRMF/DICE in Equation~\eqref{eq:mf}. We tune the smooth hyper-parameter $\Tilde{\gamma}$.
%This method is the modification of the BPRMF/DICE algorithms. We substitute $f_{\Theta}(u,i)$ with the  output of BPRMF/DICE in Equation~\eqref{eq:model} and then fine-tune the $\gamma$.

% \noindent $\bullet$ \textit{Hyper-parameters and Metrics.}
\subsubsection{Hyper-parameters and Metrics.}
For a fair comparison, all methods are optimized by BPR loss and tuned on the validation data. We optimize all models with the Adam~\cite{adam} optimizer with batch size as $2,048$ and default choice of learning rate (\ie 0.001) for all the experiments. We search $L_2$ regularization coefficients in the range of $\{0.01, 0.001\dots,10^{-6},0\}$ for all models. % The learning rate is fixed to $0.001$ which is the default choice of Adam~\cite{adam}.
We adopt the early stopping strategy that stops training if Recall@20 on the validation data does not increase for 100 epochs. For DICE, we use its provided stopping strategy and tune the threshold for stopping to make it comparable to our strategy.
%%% learning rate based on BPRMF
For PD and PDA, we search $\gamma$ in $[0.02,0.25]$ with a step size of $0.02$. For BPRMF-A and DICE-A, we search $\Tilde{\gamma}$ from $0.02$ to $1$ with a step size of $0.02$ and stop searching if the evaluation results do not change for $3$ steps. 
%For all MF-based methods, we randomly pair one non-clicked item as the negative sample for each interaction during training. For DICE, we keep its setting as suggested in the original paper~\cite{DICE}.
%%% 对于每个交互而言,我们直接选择一个用户未交互过的item作为负样本
%%% 我重写了下

% To measure recommendation performance, we conduct top-$K$ recommendations and adopt a full ranking strategy. We report the results with regard to four widely-used evaluation metrics: Recall@K, Precision@K,  HitRate@K(HR@K), NDCG@K. We vary  $K$ from 20 to 50.
To measure the recommendation performance, we adopt four widely-used evaluation metrics: Hit Ratio (HR), Recall,  Precision, which consider whether the relevant items are retrieved within the top-$K$ positions, and NDCG that measures the relative orders among positive and negative items in the top-$K$ list. All metrics are computed by the all-ranking protocol --- all items that are not interacted by a user are the candidates. We report the results of $K=20$ and $K=50$ in our experiments.

\subsection{Deconfounding Performance (RQ1)}
In this section, we will first study the recommendation performance of our proposed algorithm PD. Then, we analyze its recommendation lists and showcase its rationality. 
At last, we study the necessity of computing item popularity at different stages in training (\ie computing local popularity).
%%% computing popularity感觉怪怪的，意思不完整吧，是不是改成leveraging popularity for xxx之类的

\subsubsection{Overall Performance} Table~\ref{tab:PD-result} shows the comparison of top-$K$ recommendation performance. From the table we can find:
\begin{itemize}[leftmargin=*]
\item The proposed PD achieves the best performance, and consistently outperforms all the baselines on all datasets. 
This verifies the effectiveness of PD, which is attributed to the deconfounded training to remove the bad effect of popularity bias. In addition, the superior performance of PD reflects the rationality of our causal analysis and the potential of \textit{do}-calculus in mitigating the bias issue in recommendation.
% with do-calculus.
%%%不确定这个PD是不是统一加公式符号，我先都加上了
\item PD achieves different degrees of improvement on the three datasets. In particular, PD outperforms all baselines by at least $119\%$, $34\%$, and $24\%$ on Kwai, Douban, and Tencent, respectively. We think the difference in the improvement is due to the different properties of the three datasets. For example, the level of popularity drift in the data differs substantially as shown in Figure~\ref{fig:drit_showing}: the popularity drifts of Kwai are much larger than that of the other two datasets. 
%Higher drifts mean more different instances of $P(C|u, i, Z)$ on Z. Normally, more varied instances on Z will help the model to estimate $P(C|u, i, Z)$ better. 
Higher drifts imply that item popularity $Z$ has larger impact on the recommendation data. 
That is why the proposed PD outperforms baseline methods the most on the Kwai dataset. Hence, we believe that formulating the recommender model as $P(C|do(U, I))$ has larger advantages over $P(C|U, I)$ in the scenarios exhibiting higher popularity drift.% For the other two datasets, training and testing data are sampled from less varied distribution. Hence, the improvements are less compared with the models that estimate $P(C|U, I)$.
%It's obvious that Kwai has more difference as showing in figure~\ref{fig:drit_showing}.

\item For BPR-PC and xQuad, which perform ranking adjustment, only BPR-PC has limited improvements over BPRMF on recall, \ie ranking adjustment has %. It seems that the two methods have 
little effect in our setting. The reason is that: heuristic adjustment of the original model prediction can only guarantee an increase in the proportion of long-tail items but cannot guarantee the promoted items are of real interest. BPR-PC performs better than xQuad since  xQuad only considers group-level popularity, while BPR-PC computes popularity in a finer granularity at item level.%. In BPR-PC, popularity for each item is considered, thus resulting in higher accuracy.

\item DICE tries to disentangle popularity and interest influence for an observed interaction to get causal embedding. It only demonstrates a similar performance to BPRMF in our setting, which indicates the rationality of leveraging popularity bias instead of blindly eliminating the bias. 
%We fail to reproduce the results in the original paper since they manually sampled even data for both training and testing while we only use real-world user interactions. 
Note that in the original paper of DICE, the evaluation data is manually intervened to ensure an even distribution of testing interactions over items, which removes popularity bias in the testing data. Whereas in our setting, the evaluation data is split by time to reflect real-world evaluation of recommender system, and contains popularity bias. 
Although DICE boosts the scores of unpopular items over popular items, it may under-estimate users' interest matching for popular items.

%\item Note that MostPop merely ranks items based on the global popularity, and produces the same ranking list for all users. It performs slightly worse than MF ($14.8\%$) in the Douban dataset, which indicates that some bias can be useful in producing recommendations.
\end{itemize}

\begin{table*}[]
\caption{Top-K recommendation performance with popularity adjusting on Kwai, Douban, and Tencent Datasets.} 
\vspace{-8pt}
\label{tab:PDA-result}
\resizebox{0.9\textwidth}{!}{%
\begin{tabular}{c|c|cccc|cccc|cccc}
\hline
\multicolumn{2}{c|}{Datasets} & \multicolumn{4}{c|}{Kwai} & \multicolumn{4}{c|}{Douban} & \multicolumn{4}{c}{Tencent} \\ \hline
\multicolumn{2}{c|}{} & \multicolumn{2}{c|}{top 20} & \multicolumn{2}{c|}{top 50} & \multicolumn{2}{c|}{top 20} & \multicolumn{2}{c|}{top 50} & \multicolumn{2}{c|}{top 20} & \multicolumn{2}{c}{top 50} \\
\multicolumn{2}{c|}{\multirow{-2}{*}{Methods}} & Recall & \multicolumn{1}{c|}{NDCG} & Recall & NDCG & Recall & \multicolumn{1}{c|}{NDCG} & Recall & NDCG & Recall & \multicolumn{1}{c|}{NDCG} & Recall & NDCG \\ \hline
\multicolumn{2}{c|}{MostRecent} & 0.0074 & 0.0096 & {\color[HTML]{0D0D0D} 0.0139} & 0.011 & 0.0398 & 0.0582 & 0.0711 & 0.0615 & 0.0360 & 0.0222 & 0.0849 & 0.0359 \\ \cline{1-2}
\multicolumn{2}{c|}{BPRMF(t)-pop} & 0.0188 & 0.0241 & 0.0372 & 0.0286 & 0.0495 & 0.0682 & {\ul 0.0929} & 0.0760 & 0.1150 & 0.0726 & 0.2082 & 0.1001 \\ \cline{1-2}
 & (a) & 0.0191 & 0.0249 & 0.0372 & 0.0292 & 0.0482 & 0.0666 & 0.0898 & 0.0744 & 0.1021 & 0.0676 & 0.1805 & 0.0905 \\
\multirow{-2}{*}{BPRMF-A} & (b) & 0.0201 & 0.0265 & 0.0387 & 0.0306 & 0.0486 & 0.0667 & 0.0901 & 0.0746 & 0.1072 & 0.0719 & 0.1886 & 0.0953 \\ \cline{1-2}
 & (a) & 0.0242 & 0.0315 & 0.0454 & 0.0363 & 0.0494 & 0.0681 & 0.0890 & 0.0736 & 0.1227 & 0.0807 & 0.2161 & 0.1081 \\
\multirow{-2}{*}{DICE-A} & (b) & 0.0245 & 0.0323 & 0.0462 & 0.0370 & 0.0494 & 0.0680 & 0.0882 & 0.0734 & 0.1249 & 0.0839 & 0.2209 & 0.1116 \\ \cline{1-2}
 & (a) & {\ul 0.0279} & {\ul 0.0352} & {\ul 0.0531} & {\ul 0.0413} & {\ul 0.0564} & \textbf{0.0746} & \textbf{0.1066} & \textbf{0.0845} & {\ul 0.1357} & {\ul 0.0873} & {\ul 0.2378} & {\ul 0.1173} \\
\multirow{-2}{*}{PDA} & (b) & \textbf{0.0288} & \textbf{0.0364} & \textbf{0.0540} & \textbf{0.0429} & {\textbf{0.0565} } & {\ul 0.0745} & {\textbf{0.1066}} & {\ul 0.0843} & \textbf{0.1398} & \textbf{0.0912} & \textbf{0.2418} & \textbf{0.1210} \\ \hline
\end{tabular}
}
\vspace{-10pt}
\end{table*}

\begin{figure}
\centering
\subfigure[\vspace{-5pt}\textbf{ Kwai}]{\includegraphics[width=0.23\textwidth]{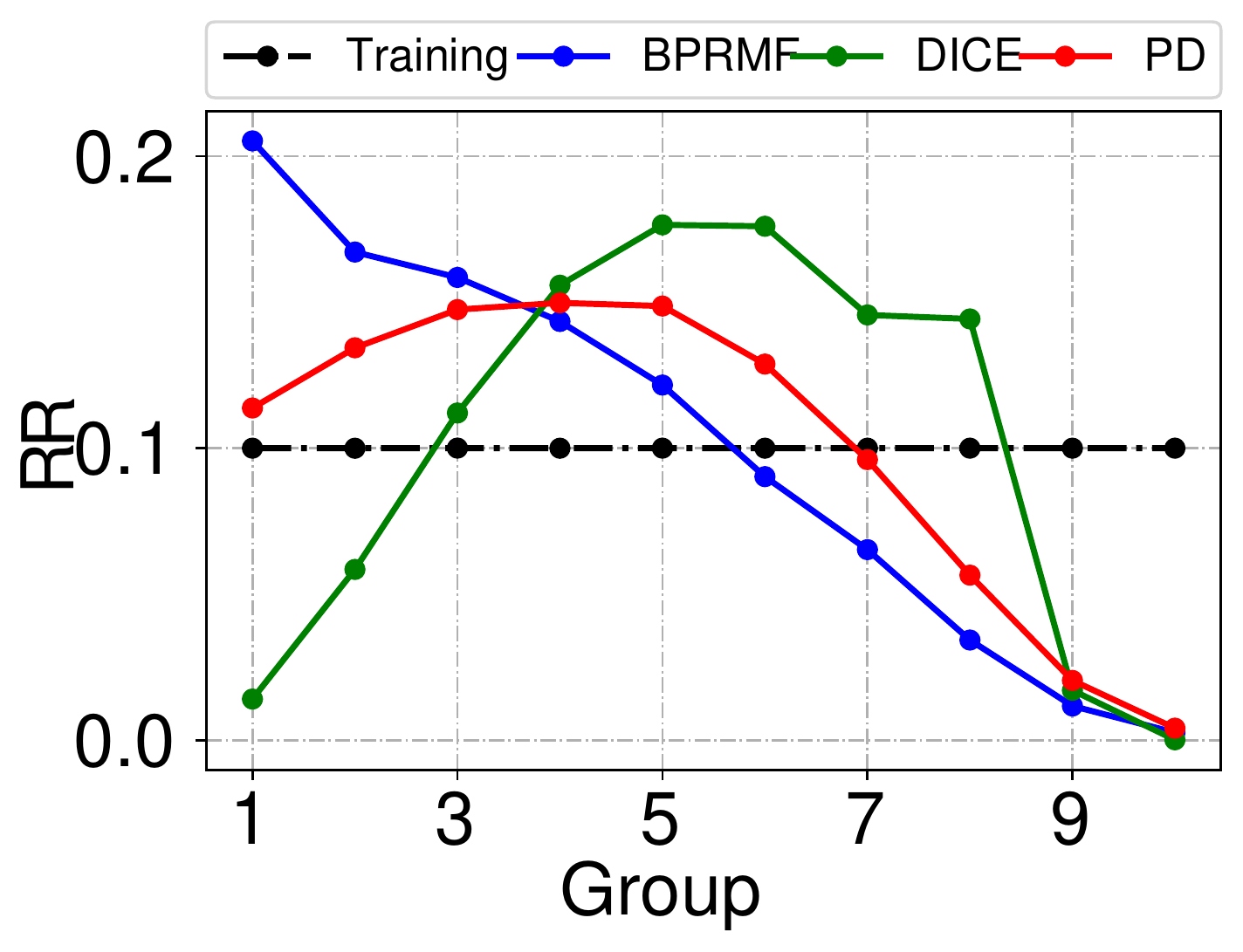}}
\subfigure[ \textbf{Douban}]{ \includegraphics[width=0.235\textwidth]{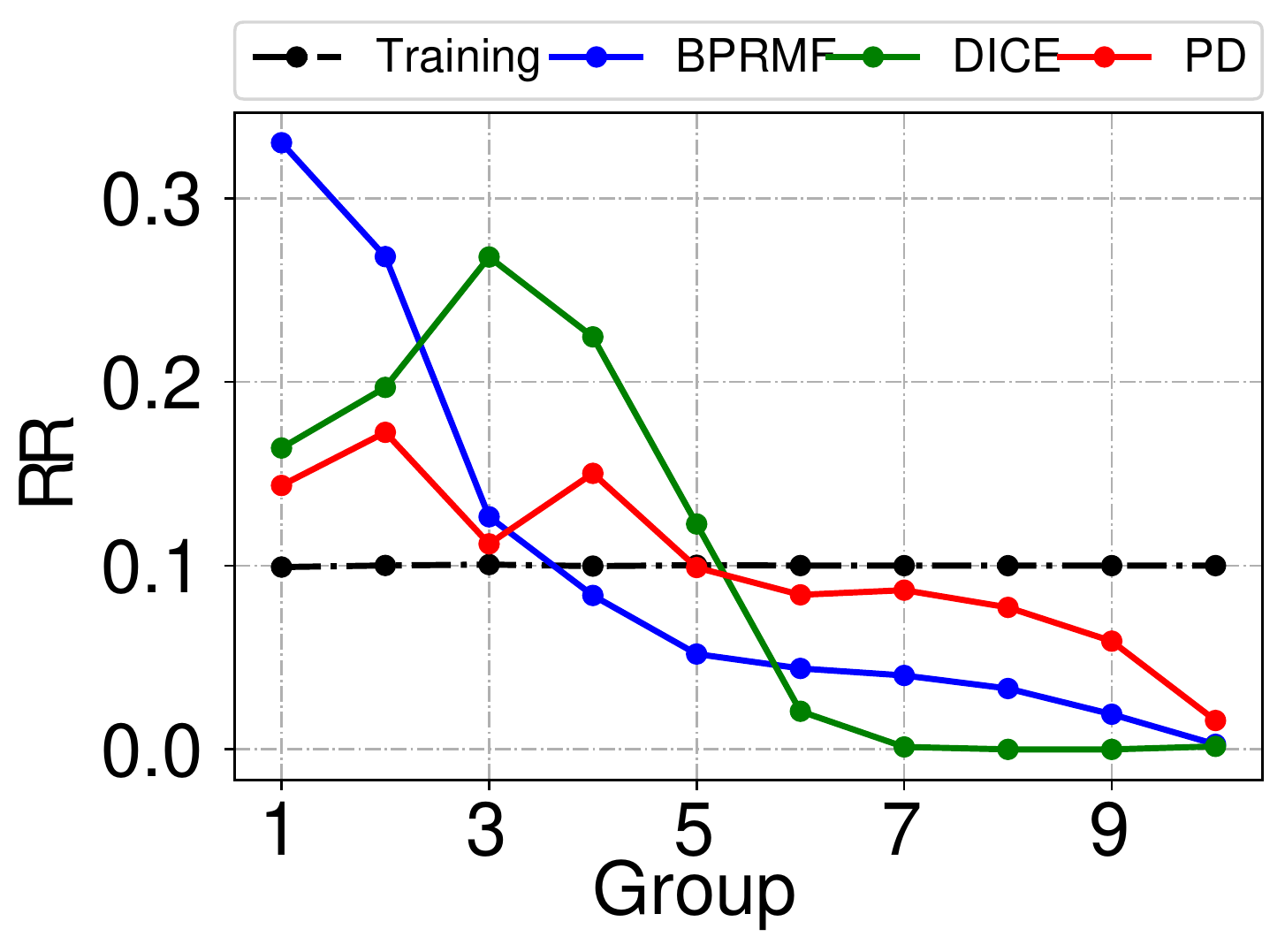}}
\quad
\subfigure[ \textbf{Tencent}]{ \includegraphics[width=0.23\textwidth]{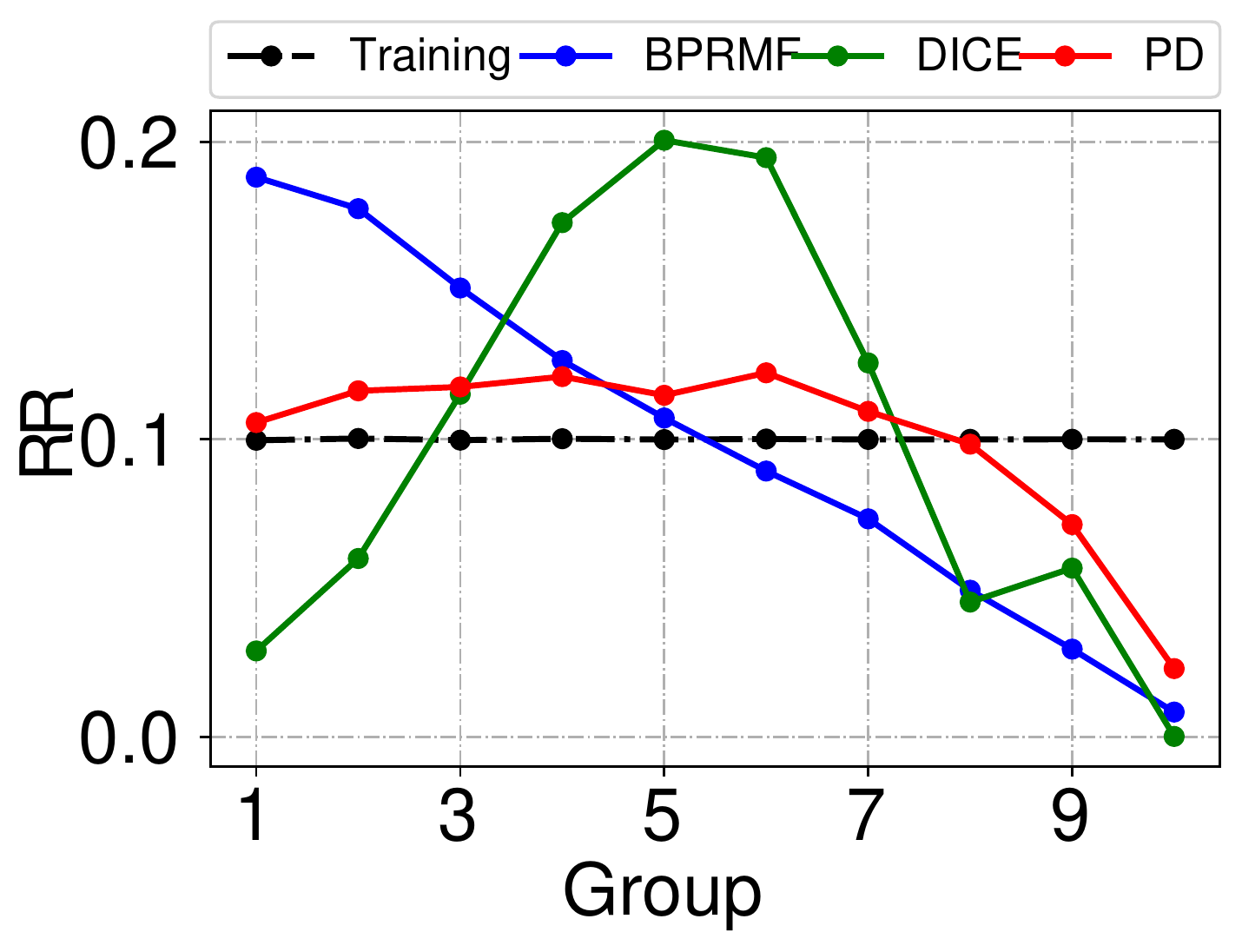}}
\subfigure[ \textbf{std. dev.}]{ \includegraphics[width=0.23\textwidth]{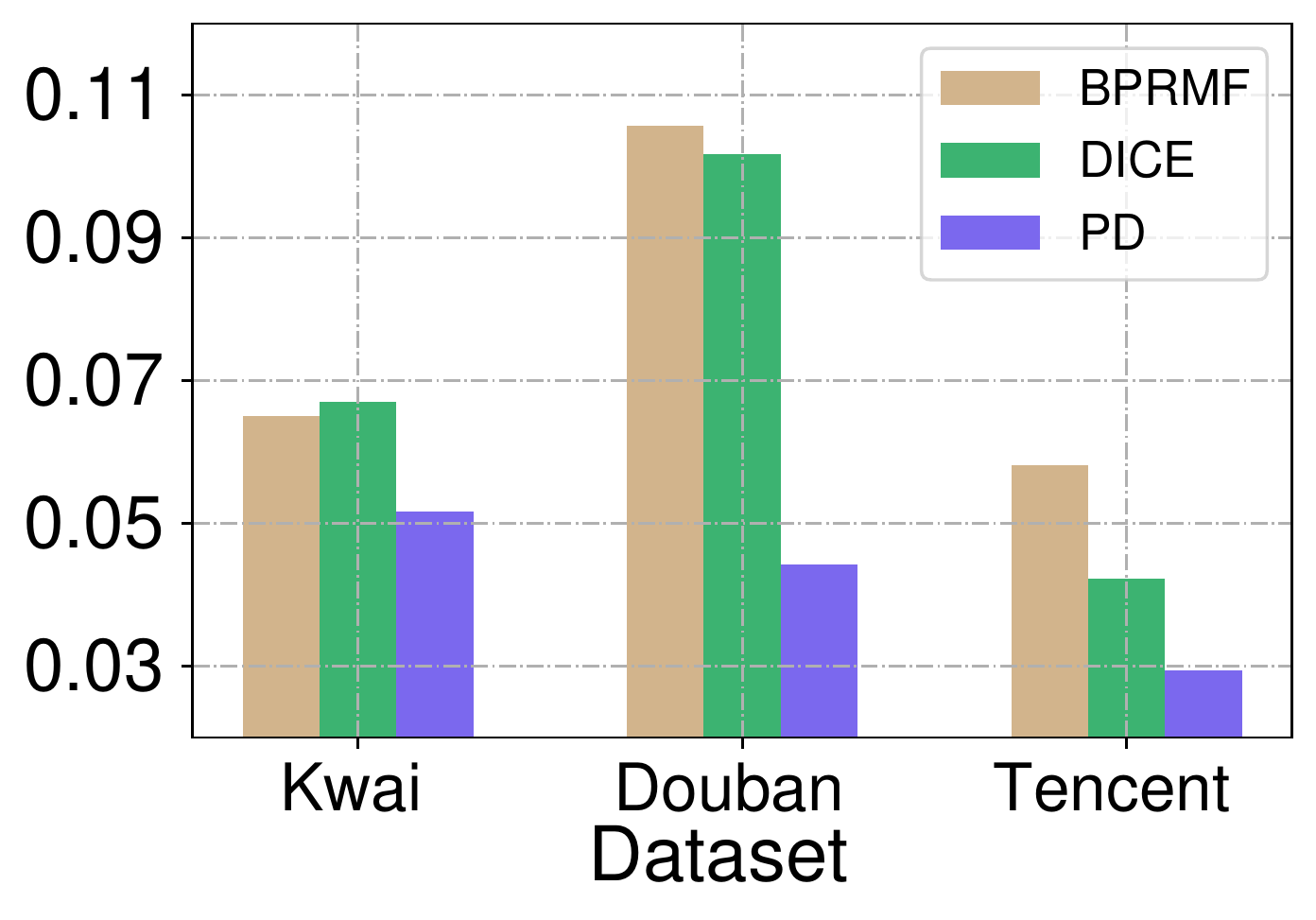}}
\vspace{-15pt}
\caption{Recommendation rate(RR) over item groups.}
\label{fig:group-rec}
\vspace{-18pt}
\end{figure}

\subsubsection{Recommendation Analysis}
We claim that PD can alleviate the amplification of popularity bias, \ie removing the bad effect of popularity bias. We conduct an analysis of the recommendation lists to show that deconfounded recommendation does not favor popular items as much as conventional recommendations. 
%We split items into 10 groups, and compare the recommendation results for different groups between different methods. 
Items are placed into $10$ different groups in two steps: (1) we sort items according to their popularity in descending order and (2) we divide the items into $10$ groups and ensure that the \emph{sum} of popularity over items in each group is the same. As such, the \emph{number} of items will increase from group 1 to group 10 and the \emph{average} popularity among items in each group decreases from group 1 to group 10. Hence we say group 1 is the most popular group and group 10 is the least popular group.

For each method, we sum the number of times of items recommended from each group. Then we divide this number by the total number of recommendations to get the recommendation rate for each group.  Formally, we define recommendation rate (RR) for group $g$ as:
\begin{equation}\nonumber \small
    RR(g,algo)= \frac{\sum_{i\in Group_g}{RC(i, algo)}}{\sum_{i\in All Groups}{RC(i,algo)}}
\end{equation}
where $RC(i, algo)$ gives the number of times the algorithm $algo$ recommends item $i$ in  top-$K$ recommendation lists.

For training, the recommendation rate of different groups is uniform (see the black line in Figure~\ref{fig:group-rec}). If more popular groups have higher recommendation rates, there is the amplification of popularity bias. We show the recommendation rates of each group by different methods in Figure~\ref{fig:group-rec}, in which (a), (b), and (c) are the results on Kwai, Douban, and Tencent, respectively. Figure~\ref{fig:group-rec} (d) is to show the standard deviation (std. dev.) of recommendation rates over all the groups: a smaller value means the recommendation rate is more uniform and is closer to the recommendation rates from training set. The main findings are as follows.
\begin{itemize}[leftmargin=*]
\item It is clear that popularity bias will be amplified by BPRMF. Figures~\ref{fig:group-rec}~(a) (b) (c) show that as the average item popularity decreases from group 1 to group 10, the recommendation rate decreases almost linearly. The gap between the chances to recommend popular items and unpopular items becomes bigger in the recommendation lists compared with that in the training set.

\item For DICE, the recommendation rate increases initially and then decreases from group 1 to group 10. This is because DICE strategically increases the scores of unpopular items to suppress the problem of amplified interest for popular items. However, this suppression brings two side effects: (1) the items from group 1 and group 2 are over-suppressed such that they have lower chances to be recommended than they have in the training set on Kwai and Tencent and (2) it over-amplifies the interest for the middle groups that contain sub-popular items. As such, blindly eliminating popularity bias would remove the beneficial patterns in the data.

\item In Figure~\ref{fig:group-rec}~(a) (b) (c), the RRs of PD exhibit lines that are flatter than other methods, which are the most similar to the training set. Moreover, as we can see in Figure~\ref{fig:group-rec}~(d), the RR of PD has the smallest standard deviation across groups, indicating the least difference across groups. These results verify that PD provides relatively fair recommendations across groups, and will not over-amplify or over-suppress interest for popular items. Along with the recommendation performance in Table~\ref{tab:PD-result}, the results demonstrate the rationality and effectiveness of only removing the bad effect of popularity bias by deconfounded training.%and meanwhile keep the good effect.

\item For the group $10$, which has the least popular items, PD has a little higher recommendation rate than other methods. However, PD also cannot provide enough chances of recommendation because of the sparse interactions of items in this group and the %. Items in group $10$ have very few interactions, resulting in 
insufficient learning of item embedding. 
\end{itemize}

Note that we omit the RRs of the ranking adjustment methods (\ie BPR-PC and xQuad) since their recommendation rates for unpopular items are manually increased. Instead of tuning the recommendation rate for groups of different popularity, we concentrate more on whether the learned embedding can precisely reveal the real interests of users.

\subsubsection{Global Popularity \textit{V.S.} Local Popularity.} 
\begin{figure}
    \centering
    \subfigure[\textbf{ Recall}]{\label{fig:a}\includegraphics[width=0.23\textwidth]{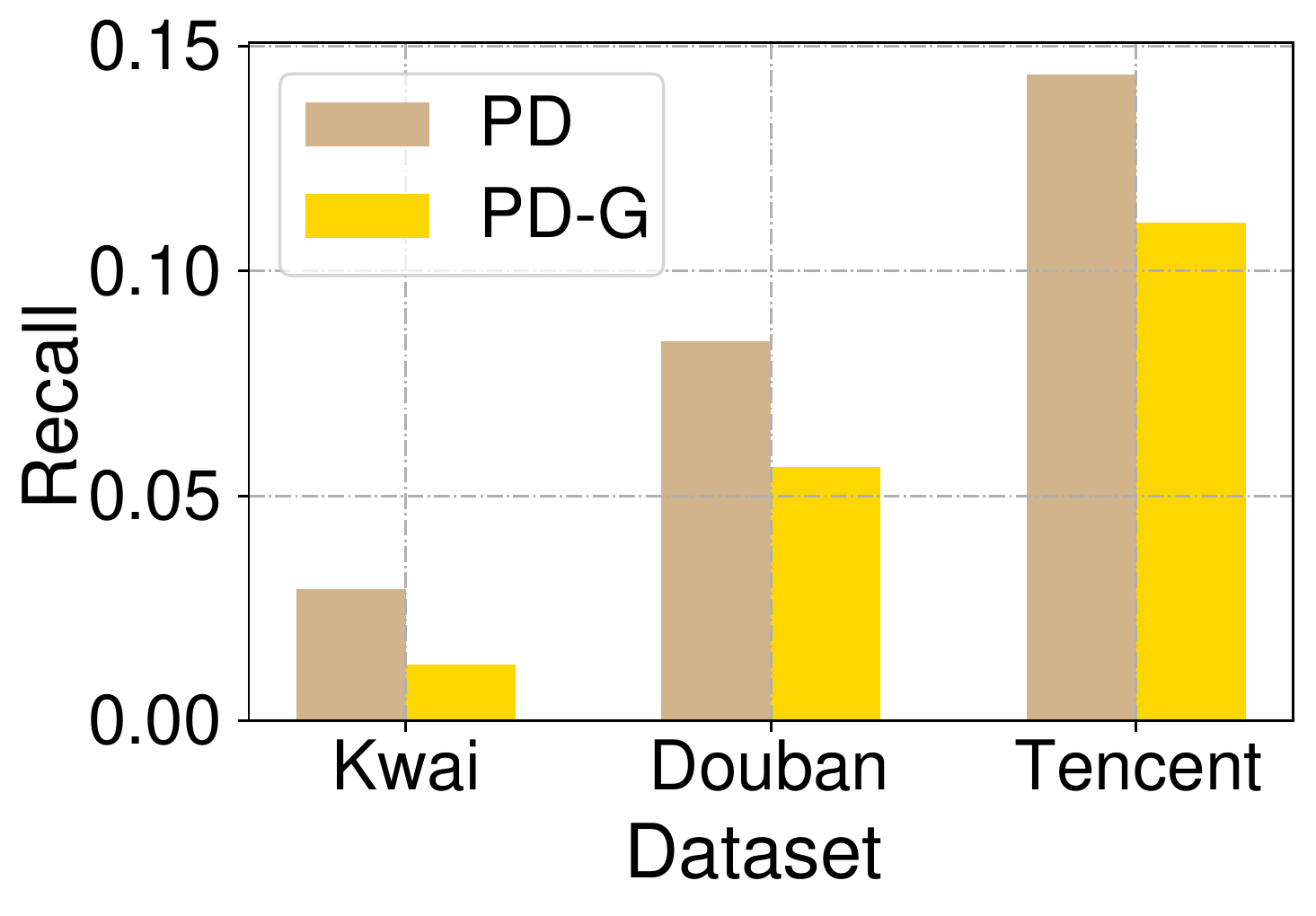}}
    \subfigure[ \textbf{NDCG}]{\label{fig:b} \includegraphics[width=0.23\textwidth]{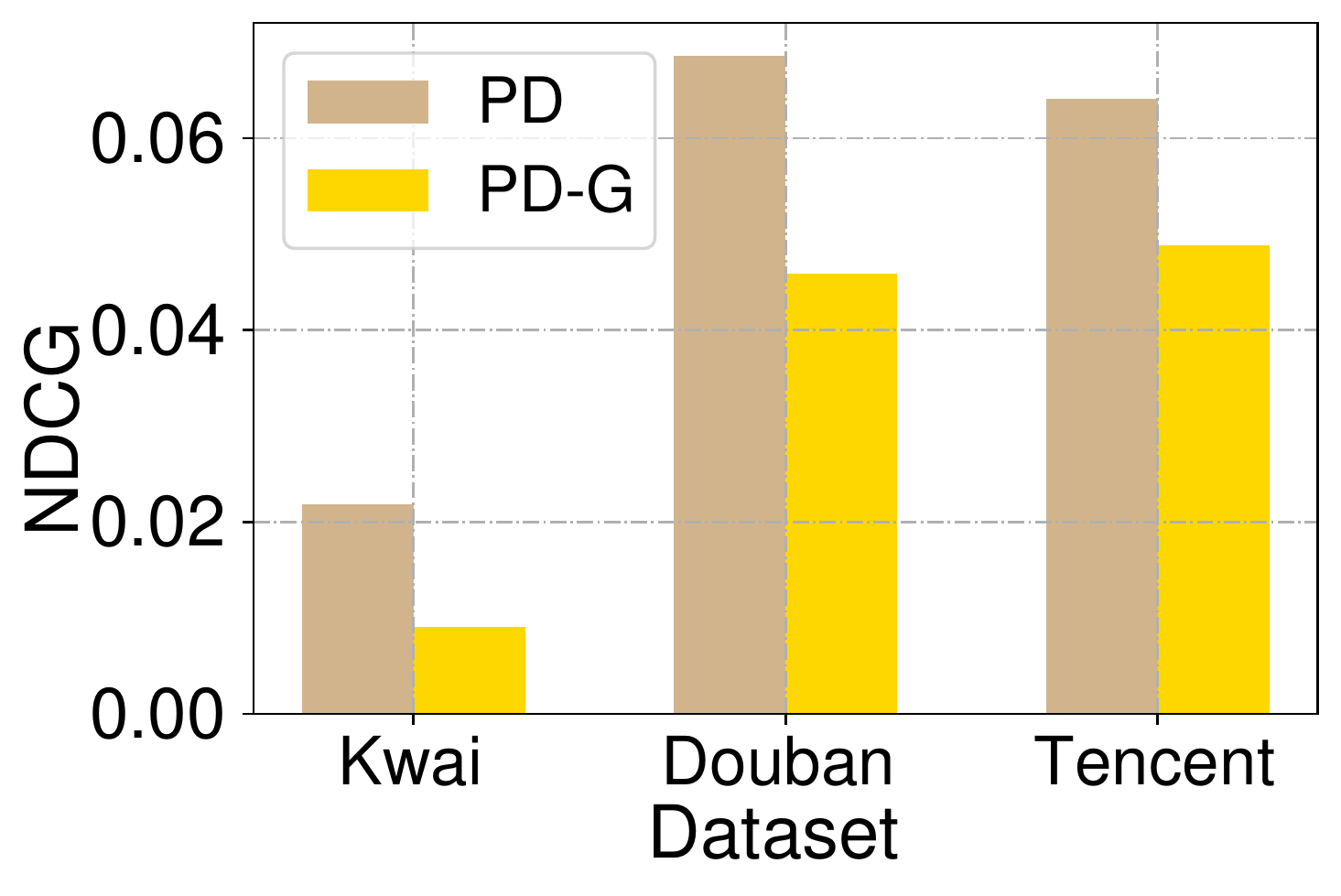}}
    \vspace{-15pt}
    \caption{Comparisons between PD and PD-G on Recall@50 and NDCG@50. PD-G is a version of PD that computes item popularity over the total training set.}
    \vspace{-15pt}
    \label{fig:PD-G}
\end{figure}

To investigate the necessity of local popularity, we evaluate a variant of PD that uses global popularity defined on the interaction frequency in the entire training set $\mathcal{D}$. We keep the other settings the same as PD and name this method as PD-Global (PD-G). Figure~\ref{fig:PD-G} shows the recommendation performance of PD and PD-G on the three datasets. Due to the limited space, we only show the results on recall and NDCG for top-$50$ recommendation, other metrics show the similar results. According to the figure, PD-G is far worse than PD, which verifies the importance of calculating item popularity locally. We believe PD-G performs poorly because there is only one instance of $P(C|u, i, Z)$ on $Z$, making $P(C|u, i, Z)$ hard to be learned well.

\subsection{Performance of Adjusting Popularity (RQ2)}
In this section, we verify the performance of our proposed method PDA, which introduces some desired popularity bias into recommendation. Firstly, we compare our method with the baselines that inject some popularity bias as well --- MostRecent, BPRMF (t)-pop, BPRMF-A, and DICE-A. Secondly, we study the influence of refined prediction of item popularity for PDA. Lastly, we demonstrate overall improvements compared with the baseline model BPRMF.

\subsubsection{Comparisons with Baselines.} We utilize two methods to predict the popularity of the testing stage. Specially, method (a) takes the popularity of items from the last stage $\mathcal{D}_T$ as the predicted popularity, while method (b) computes populairity as Equation~(\ref{eq:pre_pop}). 
We apply both (a) and (b) to PDA, DICE-A, and BPRMF-A. The results are shown in Table~\ref{tab:PDA-result}, which lead to the following observations:
\begin{itemize}[leftmargin=*]
\item By comparing the results of PDA, DICE-A, BPRMF-A, and MostRecent with the results of PD, DICE, BPRMF, and MostPop in Table~\ref{tab:PD-result}, we find that introducing desired popularity bias into recommendations can improve the recommendation performance. This validates the rationality of leveraging popularity bias, \ie  promoting the items that have the potential to be popular (desired popularity bias) benefits recommendation performance.

\item Compared with other methods, PDA can outperform both the method BPRMF(t)-pop, which is specially designed for utilizing temporal popularity, and the modified methods BPRMF-A and DICE-A. This is attributed to PDA introducing the desired popularity bias by an intervention, preventing the learning model to amplify the effect of popularity and injecting desired popularity bias in inference, whereas other models are influenced by popularity bias in training.
 The results validate the advantage of causal intervention in more accurately estimating true interests.
\item On Kwai and Tencent, method (b) outperforms method (a), indicating that introducing the linear predicted popularity is more effective than straightforwardly utilizing the popularity from the previous stage; On Douban, these two methods achieve almost the same performance. This is because the Douban dataset spans over a longer period of time than the other two datasets, making it harder to predict the popularity from the trends of popularity in training stages.
\end{itemize}

\begin{figure}
    \centering
    \includegraphics[width=0.49\textwidth]{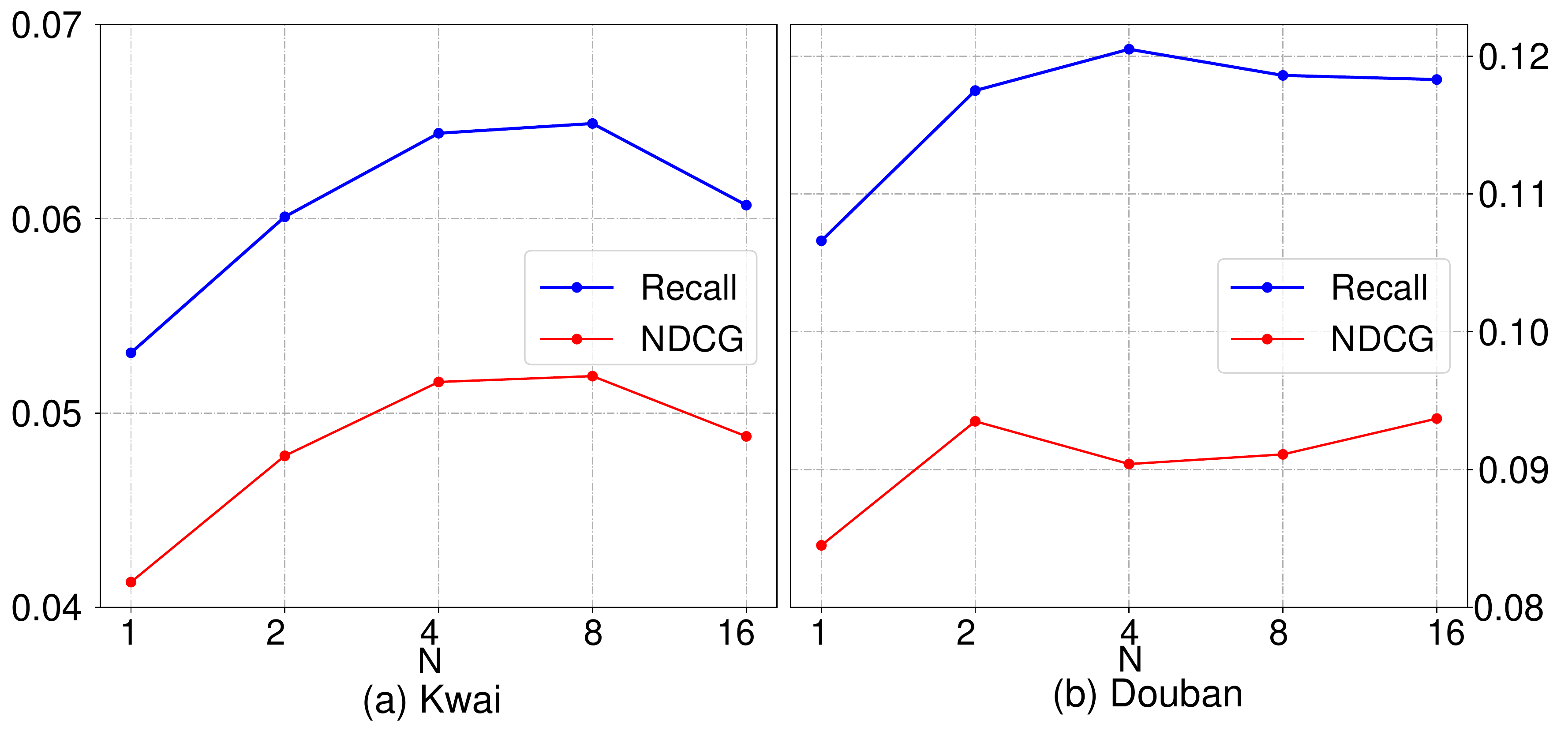}
    \vspace{-15pt}
    \caption{PDA recommendation performance regarding Recall@50 and NDCG@50 on Kwai and Douban, with the last training stage  split into $N$ sub-stages. }
    \vspace{-15pt}
    \label{fig:PD_sub}
\end{figure}

\subsubsection{More Refined Predictions for Popularity.} We conduct an experiment to show that the performance of PDA can be improved with more refined predicted popularity. 
To get more refined predictions, we uniformly split the data in the last training stage into $N$ sub-stages by time, and achieve a more refined prediction of popularity in the last sub-stage, which is closest to the testing stage. We set $N=1,2,4,8,16$ and employ method (a) to predict the desired popularity. The results are shown in Figure~\ref{fig:PD_sub}. The performance of PDA can be improved with better-predicted popularity for testing. On the Kwai dataset, the performance first increases and then slightly decreases as $N$ increases. However, on the Douban dataset, with a larger $N$, the performance does not drop evidently. 
This is mainly because the item space of Kwai is far larger than Douban, but they have a similar number of interactions. When $N$ is set to a large value, the interaction frequency is insufficient to accurately compute item popularity for Kwai. This experiment shows that the performance of PDA can be further improved with a more precise prediction of popularity.

\subsubsection{Total Improvements of PDA} To quantify the strengths of PDA, we show the relative improvements of PDA over the basic model that PDA is built on, \ie BPRMF in Table~\ref{tab:total-improvement}. On the Kwai dataset, we get more than $241\%$ improvements over all metrics. On the Douban and Tencent datasets, for the recall and NDCG, we get at least $97\%$ improvements, and for precision and HR, we get at least $57\%$ improvements. The results clearly indicate the superiority of PDA in recommendation and the huge potential of causal analysis and inference for leveraging bias in recommendation.

\section{Related Works}
%In this section, we introduce the related works regarding the following two topics:  popularity bias in recommender systems  and causal recommendation. 
%There are two series of works that related to our work. The first is regarding bias. The second is to utilizing popularity and popularity drifting.
This work is closely related to two topics: popularity bias in recommender system and causal recommendation. 
\begin{table}[]
\caption{The total relative improvements of PDA compared with the base recommendation model, \ie BPRMF. Each result is averaged on top 20 and top 50.}
\vspace{-10pt}
\label{tab:total-improvement}
\resizebox{0.4\textwidth}{!}{%
\begin{tabular}{c|cccc}
\toprule
Dataset & Recall & Precision & HR & NDCG \\ \hline
Kwai & 488\% & 441\% & 241\% & 532\% \\
Douban & 124\% & 68\% & 57\% & 97\% \\
Tencent & 133\% & 133\% & 79\% & 161\% \\ \hline
\end{tabular}
}
\vspace{-15pt}
%\vspace{-15pt}
% \vspace{-0.5cm}
\end{table}

\subsection{Popularity Bias}
Popularity bias has long been recognized as an important issue in recommendation systems.
%But they focus on different aspects of popularity. 
One line of research~\cite{connection2020,UMAP_recbias,FollowCrowd,FeedbackLoop,PopAcc} studies the phenomenon of popularity bias. %and how to handle it.  
\cite{UMAP_recbias} studies how popularity bias influences  different recommendation models and concludes that popularity has a larger impact on BPR-based models \cite{DICE} than point-wise models \cite{ncf}; \cite{connection2020} tries to study the relation between popularity bias and other problems such as fairness and calibration.
Another line of research~\cite{DICE,BPR_PC,xquad2019,Himan_reg,Himan-thesis,IPS_rec,2018causalE} attempts to improve  recommendation performance by removing popularity bias. These works mainly consider three types of methods: IPS-based methods~\cite{IPS_rec,rec_ips_cap_norm}, causal embedding-based methods~\cite{2018causalE,distillation,DICE}, and the methods based on post-hoc re-ranking or model regularization~\cite{BPR_PC,xquad2019,Himan_reg,correcting-pop-bias,2011-PPT}.
Similar to our work, a very recent work \cite{DICE}  also tries to compute the influence of popularity via causal approaches. 
The difference is that we analyze the causal relations from a confounding perspective whearas this important aspect was not considered in their work. 
%Further, we solve the confounding problem by XXXXX condition on popularity while they model popularity   lies in how we model true user interest in an given item.  They  disentangle the users and items into two sets of embedding,\ie let the model  learn the popularity, and estimate $P(C|U,I)$ whereas our work precisely identifies the causal relations between user, item and popularity and estimates  $P(C|U,I,M)$ instead.
Some other works have attempted to remove the impact of other bias problems~\cite{chen_survey},  position bias~\cite{position-bias}, selection bias~\cite{wang2019doubly}, and exposure bias~\cite{exposure-modeling,chen2018modeling} \textit{etc}. The most widely considered method for handling aforementioned biases is IPS-based method~\cite{IPS_rec,ips_cjl,unbiased_PS,postclick-ips}. %Following the study of IPS, some works focus on estimating better propensity scores~\cite{ips_cjl,unbiased_PS}. 
However, there is no  guaranteed accuracy for estimating propensity scores. Utilizing unbiased data is another option for solving bias problems~\cite{2018causalE,distillation},   however the unbiased data is difficult to obtain and may hurt user experience~\cite{wang2019doubly}. 

%Doubly robust models~\cite{double_unbias_data,wang2019doubly}, which improve IPS-based method by generating data via imputation models are also  utilized for solving bias problems. 
%We don't find it has been taken to solve popularity bias. 
%Some other work try to model the influence of factors that cause bias, such as \cite{exposure-modeling} models exposure in their model as a hidden variable. 
%$Re-ranking$ methods are also considered, such as \cite{sigir_best2020}.

Some other  methods recognize the importance of utilizing popularity and consider    drifted popularity to improve the model performance~\cite{temp_pop2017,MostRecent,POP_music,ECIR-localPop}.  %\cite{FollowCrowd} identifies the condition that  popularity can be potentially utilized for better recommendation performance.
\cite{POP_music} tries to utilize temporal popularity  in music recommendation and \cite{POP-news}  utilizes local popularity for recommending  news. \cite{temp_pop2017,ECIR-localPop,MostRecent}  also try to utilize temporal popularity for general recommendation. The biggest difference between our work and other works is that we precisely differentiate popularity bias to be removed and popularity drift to be taken advantage of.
\vspace{-8pt}
\subsection{Causal Recommendation}
Some efforts have considered causality in recommendation. The first type of work is on confounding effects~\cite{CauInferRec,CE_for_rec,alg-confounding,deconfounding-response-rate}.  For example, \cite{CauInferRec}  takes the  de-confounding technique in linear models to learn real interest influenced by unobserved confounders. They learn substitutes for unobserved confounders by fitting exposure data. 
\cite{CE_for_rec} tries  to estimate the causal effect of the  recommending operation. They consider the features of users and items as confounders, and reweight training samples to solve the confounding problem. 
\cite{alg-confounding} explores the impact of algorithmic confounding on simulated recommender systems. \cite{deconfounding-response-rate} identifies  expose rate as a confounder for user satisfaction estimation and uses IPS to handle the confounding problem. 
%%%% 前面有工作利用ips技术来解决这个问题，如\cite{CE_for_rec}。
%%%跟前面工作的差异不是很清晰。 与他们不同的是，我们对于popularity bias问题来考虑confounder的影响，问题不同。
In this work, we introduce a different type of confounder that is brought by item popularity.

The second type is counterfactual learning methods~\cite{counterfactual-music,double_unbias_data,CVIB}. \cite{counterfactual-music} utilizes a quasi-experimental Bayesian framework to quantify the effect of treatments on an outcome to generate counterfactual data, where counterfactual data refers to the outcomes of unobserved treatments.
%which means not really existed data to compute the causal impact of an intervention (a new track release).
\cite{double_unbias_data} utilizes a model that is learned with few unbiased data to generate labels for unobserved data. 
%which will help the recommendation model learning. 
\cite{CVIB} extends  Information Bottleneck~\cite{CVIB} to counterfactual learning with the idea that the counterfactual world should be as informative as the factual world to deal with the  Missing-Not-At-Random problem~\cite{wang2019doubly}.  
Though these works study the problem of causal recommendation, they do not specifically consider popularity bias as well as popularity drift for boosting recommendation performance.

%补充我们的差异：与他们不同，我们不利用反事实的思想或者技术去学习模型，只是在推荐的时候有一定程度的体现。

% At last, in \cite{stable_graph} Stable learning is used to to learn a stable graph of the data generation.Stable learning not usually

%Methods that have been long studied and introduced above are  IPS-based %method~s\cite{ips_cjl,attributePS} and Doubly robust %models~\cite{double_unbias_data,wang2019doublypearl_primer}
%Besides IPS, there are more thing about causality are introducing into recommendation. 

% The other type is the work based on stable

% learning\cite{CE_for_rec} novely to estimate the causal effect of the recommending operation. In this work, we solve the popularity bias from a causal perspective. \cite{cuipeng-counterfactual} propose a method by a re-weighting according to causality can be used for bundle recommendation. 
\vspace{-5pt}
\section{conclusion}
In this work, we study how to model and leverage popularity bias in recommender system. By analyzing the recommendation generation process with the causal graph, we find that item popularity is a confounder that misleads the estimation of $P(C|U,I)$ as done by traditional recommendation models.
%, showing as the amplified interest for popular items. 
Different from existing work that eliminates popularity bias, we point out that some popularity bias should be decoupled for better usage.
We propose a Popularity-bias Deconfounding and Adjusting  framework to deconfound and leverage popularity bias for better recommendation. 
%It has two parts:(1) PD, a deconfounding framework to estimate $P(C|do(U,I))$ instead of $P(C|do(U,I))$ to remove confounding effect. (2) PDA, a method that can introduce some desired popularity bias by intervening. 
We conduct experiments on three real-world datasets, providing insightful analyses on the rationality of PDA. 

%Currently, we implement PDA with a very sample model--- MF.
This work showcased the limitation of pure data-driven models for recommendation, despite their dominant role in recommendation research and industry. We express the prior knowledge of how the data is generated for recommendation training as a causal graph, performing deconfounded training and intervened inference based on the causal graph. 
We believe this new paradigm is general for addressing the bias issues in recommendation and search systems, and will extend its impact to jointly consider more factors, such as position bias and selection bias. 
Moreover, we are interested in extending our method to the state-of-the-art graph-based recommendation models and incorporate the content features of users and items. %Furthermore, we will jointly consider more factors, such as position bias and selection bias, that could influence the data generation process rather than considering popularity bias alone. Lastly, we plan to model the content features of users and items in the causal graph. 
\begin{acks}
This work is supported by the National Natural Science Foundation of China (U19A2079, 61972372) and National Key Research and Development Program of China (2020AAA0106000).
\end{acks}
\bibliographystyle{ACM-Reference-Format}
\bibliography{0_main}
\end{spacing}
\end{document}